\documentclass[a4paper,11pt]{article}

\pdfoutput=1
\usepackage{comment}
\usepackage{fixltx2e}
\usepackage[table,xcdraw]{xcolor}
\usepackage{jcappub} 
\usepackage{float} 
\usepackage[scr=boondox]{mathalfa}
\usepackage{bm}
\usepackage{stmaryrd}
\usepackage{bbold}
\usepackage[normalem]{ulem}
\usepackage{amsmath}
\usepackage{bbm}

\def\be{\begin{equation}}
\def\ee{\end{equation}}
\def\bea{\begin{eqnarray}}
\def\eea{\end{eqnarray}}
\newcommand{\vs}{\nonumber\\}
\def\ba#1\ea{\begin{align}#1\end{align}}
\def\bg#1\eg{\begin{gather}#1\end{gather}}

\def\Msunh{\,h^{-1}\,M_\odot}

\newcommand{\refeq}[1]{Eq.~(\ref{eq:#1})}

\newcommand{\reffig}[1]{Fig.~\ref{fig:#1}}

\newcommand{\reftab}[1]{Tab.~\ref{tab:#1}}          
\newcommand{\refsec}[1]{Sec.~\ref{sec:#1}}          
\newcommand{\refapp}[1]{App.~\ref{app:#1}}

\def\Plin{P_{\rm L}}

\renewcommand{\v}[1]{\bm{#1}}

\renewcommand{\emph}[1]{\textit{#1}}

\newcommand{\vk}{\v{k}}
\newcommand{\vq}{\v{q}}

\renewcommand{\d}{\delta}

\newcommand{\G}{\mathcal{G}}

\def\L{\Lambda}

\def\L{\Lambda}

\def\kmax{k_\text{max}}

\def\P{\mathcal{P}}
\def\O{\mathcal{O}}

\newcommand{\perm}[1]{ \expandafter\ifstrempty\expandafter{#1} {\mbox{perm.}} {\mbox{$#1$ perm.}} }

\makeatletter
\newlength{\apb@width}
\newcommand{\autoparbox}[2][c]{\settowidth{\apb@width}{#2}\parbox[#1]{\apb@width}{#2}}

\makeatother

\DeclareMathOperator{\Cov}{Cov}

\renewcommand{\comment}[1]{}

\usepackage[T1]{fontenc}
\usepackage{subfig}
\usepackage{booktabs}
\title{\boldmath Multi-tracer beyond linear theory}

\author[a,b,c]{Henrique Rubira}
\author[a]{Francesco Conteddu}

\affiliation[a]{University Observatory, Faculty of Physics, Ludwig-Maximilians-Universit\"at, Scheinerstr. 1, D-81679 München, Germany}
\affiliation[b]{Kavli Institute for Cosmology Cambridge, Madingley Road, Cambridge CB3 0HA, UK}
\affiliation[c]{Centre for Theoretical Cosmology, Department of Applied Mathematics and Theoretical Physics
University of Cambridge, Wilberforce Road, Cambridge, CB3 0WA, UK}

\emailAdd{henrique.rubira@lmu.de, fconteddu@mpa-garching.mpg.de}

\abstract{
The multi-tracer (MT) technique has been shown to outperform single-tracer analyses in the context of galaxy clustering. In this paper, we conduct a series of Fisher analyses to further explore MT information gains within the framework of non-linear bias expansion. We examine how MT performance depends on the bias parameters of the subtracers, showing that directly splitting the non-linear bias generally leads to smaller error bars in $A_s$, $h$, and $\omega_{\rm cdm}$ compared to a simple split in $b_1$. This finding opens the door to identifying subsample splits that do not necessarily rely on very distinct linear biases. We discuss different total and subtracer number density scenarios, as well as the possibility of splitting into more than two tracers. Additionally, we consider how different Fingers-of-God suppression scales for the subsamples can be translated into different $k_{\rm max}$ values. Finally, we present forecasts for ongoing and future galaxy surveys.
	}


\begin{document}

\maketitle
\flushbottom

\section{Introduction}
\label{sec:introduction}

In light of the large datasets provided by galaxy surveys \cite{BOSS:2016wmc,Abbetal,eBOSS:2019dcv, DESI:2016fyo,Amendola:2012ys,Ivezic:2008fe,SPHEREx:2014bgr,PFSTeam:2012fqu}, how can we maximize the amount of information extracted from them? Attempts to extract additional information include using alternative statistics \cite{Valogiannis:2023mxf,Eickenberg:2022qvy,Rubira:2020inb,Philcox:2020fqx,Neyrinck:2009fs,Biagetti:2022qjl,Banerjee:2020umh,Seljak:2012tp} or considering field-level analysis \cite{Nguyen:2024yth}.
In this work, we focus on the multi-tracer (MT) approach as a tool to extract more information from galaxy clustering. The multi-tracer method has been extensively discussed as a way to improve the measurement of primordial non-Gaussianities and redshift-space distortions \cite{Seljak:2008xr, McDonald:2008sh, Bernstein:2011ju, Cai:2011wj, Abramo:2015iga, Abramo:2022qir, LoVerde:2016ahu, Hamaus:2012ap, Barreira:2023rxn,Karagiannis:2023lsj,Blake:2013nif,Ross:2013vla,Beutler:2015tla,Marin:2015ula,Zhang:2021uyp,Sullivan:2023qjr,eBOSS:2020xwt,eBOSS:2020rpt,eBOSS:2021pff,Chisari:2016xki, Montano:2024xrr, Tanidis:2020byi, Gomes:2019ejy, Ferramacho:2014pua, Witzemann:2018cdx,Abramo:2021irg, Boschetti:2020fxr}. Moreover, MT serves as a central tool for the SPHEREx collaboration in improving their results on primordial non-Gaussianities \cite{SPHEREx:2014bgr,Heinrich:2023qaa}.
 
Most MT analyses have been based on linear theory and the cosmic-variance cancellation argument, finding that multi-tracer techniques generally perform better when there is a larger difference in their linear bias parameters. This approach relies on the non-trivial task of identifying two samples that have some redshift overlap while having different linear bias values.
The references \cite{Mergulhao:2021kip,Mergulhao:2023zso} (see also \cite{Zhao:2023ebp,Ebina:2024ojt}) have extended the former analysis to include non-linear scales by considering additional bias coefficients and loop calculations within the effective field theory (EFT) of large-scale structure framework \cite{Baumann:2010tm, Carrasco:2012cv, Carrasco:2013mua, Konstandin:2019bay,Angulo:2015eqa,Baldauf:2021zlt,BOSS:2013eso,DAmico:2019fhj,Ivanov:2019hqk,Colas:2019ret, Philcox:2020vvt, Nishimichi:2020tvu, eBOSS:2021poy} and the large-scale bias expansion \cite{Desjacques:2016bnm}. Their work shows that multi-tracer improvements can extend beyond cosmic variance cancellation: the correlation matrix of the MT basis is more diagonal compared to that of single-tracer (ST) analyses, effectively breaking degeneracies between different bias coefficients and cosmological parameters.

In this work, we extend the study of \cite{Mergulhao:2021kip,Mergulhao:2023zso} conducing a series of Fisher analysis that allows us to easily explore different MT scenarios in the context of non-linear galaxy clustering. Instead of splitting the samples via a feature such as color or mass, we study the direct dependence of the MT gains as a function of the direct difference in their (linear and non-linear) bias parameters. 
We find that detecting samples with distinct non-linear bias coefficients leads to comparable (and often better) results than a simple split in the linear bias. Using separate-Universe based relations, we investigate the possibility of finding tracers with different non-linear bias parameter. Despite finding that it is relatively difficult to have tracers with different non-linear bias coefficients in the most vanilla scenario, it is shown that assembly bias can help to find tracers with very distinct $b_{\G_2}$ \cite{Lazeyras:2021dar}. Next, we study the information budget encoded in MT at both linear and non-linear orders. 
In the non-linear scenario, the cross-spectra adds an important piece of information compared to the linear MT.
We also find the linear case to be way more dependent on the assumption of no cross-stochasticity between the tracers. When considering the EFT modeling, the MT leads to significant better results even when including the cross-stochastic contribution, in tandem with the results of \cite{Mergulhao:2021kip,Mergulhao:2023zso}. 

Moreover, we consider for the first time a MT non-linear galaxy clustering analysis with more than two tracers. We show that, for realistic survey specifications, two tracers is the optimal number of tracers. 
Furthermore, we study the dependence of the MT information gain on the total tracer number density. We find that, while the MT gains are limited to the very high number density case, when modeling with non-linear bias coefficients MT overtakes ST already for number densities of $10^{-4} h^3$Mpc$^{-3}$.  
We also consider a non-balanced split between the two subsamples, in which one of the samples is denser compared to the other. We find that most of the MT gains are present even in the case in which one of the samples correspond to $\sim10\%$ of the total sample. This result makes easier the task of finding two samples with different bias parameters, since one can restrict to the $10\%$ less homogeneous subsample. We also discuss how MT can be used to select subsamples with different Fingers-of-God (FoG) suppression scale, such that one could use different scale cuts for different tracers. Finally, we forecast the sensitivity of current and future galaxy surveys.   

The structure of this paper is the following. In \refsec{theory} we construct discuss the bias expansion for multi-tracer and the details of the Fisher analysis performed in this work. We present the results in terms of the bias split in \refsec{biassplit}. Next, in \refsec{results} we discuss the optimal number of tracers and the dependence on the total and subtracer number density. \refsec{fog} develops on the FoG suppression for the different subsample. We present forecasts for distinct galaxy surveys in \refsec{forecasts} and conclude in \refsec{conclusion}. We have dedicated appendices for discussing the stability of the Fisher derivatives in \refapp{derivatives}, the dependence on the fiducial bias choice in \refapp{fiducial} and for extra plots in \refapp{extraplots}.

\section{Prerequisites}
\label{sec:theory}
In this section, we review the large-scale bias expansion and its extension to multi-tracer in \refsec{EFT}. Next, we describe the Fisher information in \refsec{fisher}. In \refsec{setup}, we present the setup for the Fisher analysis.

\subsection{The large-scale bias expansion} \label{sec:EFT}

We start by considering a tracer $T$ with density $\rho^T(\boldsymbol{x}, z)$ and an average background density $\bar {\rho}^T(z)$ in the redshift $z$. The bias expansion \cite{Desjacques:2016bnm} consists of expanding its overdensity 
\be
\label{eq:bias_expansion}
    \delta^{T}(\boldsymbol{x}, z)= \frac{\rho^T(\boldsymbol{x}, z)}{\bar{\rho}^T(z)} -1= \sum_{\mathcal{O}} b_{\mathcal{O}}^T(z) \mathcal{O}(\boldsymbol{x}, z) + \epsilon^T(\boldsymbol{x}, z) \,,
\ee
in terms of the most general basis of operators
\be
\label{eq:operator_basis}
    \mathcal{O} \in
    \left\{
    \delta, \delta^{2}, \mathcal{G}_{2}[\Phi_{g}] , \Gamma_{3}[\Phi_{g}, \Phi_{v}], \nabla^2 \d \right\}\,,
\ee
and their respective bias parameters $b_{\mathcal{O}}^T(z)$, with $b_1 = b_\delta$. 
These operators are constructed as functions of the gravitational and velocity potentials $\Phi_{g}$ and $\Phi_{v}$, respectively. 
In this work, we consider only the operators in \refeq{operator_basis} that contribute to the one-loop power spectrum.\footnote{Notice that other third-order operators, such as $\delta^{3}, \delta\mathcal{G}_{2}[\Phi_{g}]$ and $\mathcal{G}_{3}[\Phi_{g}]$, in principle, contribute to the one-loop power spectrum, but they can be removed by renormalization \cite{Chudaykin:2020aoj}.} Another central element in the bias expansion is the presence of a stochastic component $\epsilon^T$ \cite{Cai:2010gz, Fang:2023ypv, Rubira:2023vzw,Rubira:2024tea}, which leads to corrections of small-scale modes onto large scales, such as the shot-noise term.

The auto power spectrum of the tracer $T$ is therefore given by
\ba
\label{eq:Pst}
 P^{TT}(k,z) =  \sum_{\O_\alpha,\O_\beta} b_{\O_\alpha}^T(z)b_{\O_\beta}^T(z)\P_{\O_\alpha,\O_\beta}(k,z) + P_{\rm c.t.}^{TT}(k,z) + P_{\rm stoch}^{TT}(k,z)  \,, 
\ea
where the functions $\P_{\O_\alpha,\O_\beta}$ can be calculated taking contractions of the operators $\O_\alpha$ and $\O_\beta$ (see e.g. \cite{Chudaykin:2020aoj} for the form of those operators). 
The counter-term to the one-loop power spectrum is given by the operator $\nabla^2 \delta$
\be \label{eq:PTT}
    P_{\rm c.t.}^{TT}(k,z) = -2b^T_{\nabla^2\delta}(z)b^T_\d(z)\frac{k^2}{k_{\rm norm}^2}\Plin(z) \,,
\ee
where $k_{\rm norm} = 0.1 \,h/$Mpc is a normalization scale chosen such that $b^T_{\nabla^2\delta}$ is dimensionless and with $\Plin$ denoting the linear matter power spectrum.
The stochastic term is calculated as the contribution of $\epsilon^T$ to large-scale modes and can be parametrized (at leading-order in perturbation theory) as a constant and a $k^2$ term
\be \label{eq:shotnoise}
    P_{\rm stoch}^{TT}(k,z) = \frac{1}{\Bar{n}_T}\left(1 + c^{TT}_0(z) + c^{TT}_2(z)\frac{k^2}{k_{\rm norm}^2}\right)\,,
\ee
where $\Bar{n}_T$ is the number density of the tracer $T$. For the linear case, we only keep the constant shot-noise term $c_0$.

The multi-tracer approach consists of splitting the tracer $T$ into (disjoint) sub-tracer samples $t_i$ based on specific criteria e.g., the mass of the sample, color, spin or the star formation rate. The total density can then be written as the sum of its parts
\be
\rho^T(\boldsymbol{x}, z) = \sum_{t_i} \rho^{t_i}(\boldsymbol{x}, z)\,, \quad \textrm{such that} \quad \delta^{T}(\boldsymbol{x}, z)  = \frac{1}{\bar{n}^T} 
    \sum_{t_i} \bar{n}^{t_i}\delta^{t_i}(\boldsymbol{x}, z) \,,
\ee
with $\bar{n}^T = \sum_{t_i} \bar{n}^{t_i}$.
The spectra $P^{t_it_j}(k,z)$ for two sub-tracers with overdensities $\delta^{t_i}(\boldsymbol{x}, z)$ and $\delta^{t_j}(\boldsymbol{x}, z)$ can be written as a generalization of \refeq{Pst}
\ba
\label{eq:generalP}
 P^{t_it_j}(k,z) =  \sum_{\O_\alpha,\O_\beta} b_{\O_\alpha}^{t_i}(z)b_{\O_\beta}^{t_j}(z)\P_{\O_\alpha,\O_\beta}(k,z) + P_{\rm c.t.}^{t_it_j}(k,z) + P_{\rm stoch}^{t_it_j}(k,z)  \,,  
\ea
which includes the auto-spectra of the tracer $t_i$ when $j=i$.
This is a fundamental point for MT: instead of considering a single species with only a single auto spectrum it breaks this species into sub-tracers also including their cross-correlations \cite{Abramo:2021irg}.
The tracer split can lead to different auto- and cross-spectrum shapes that are otherwise averaged out in the single-tracer auto spectrum.
For example, one can see stronger Fingers-of-God (FoG) effects in one of the samples \cite{Mergulhao:2023zso}.
Altogether, this can help to break the degeneracy between bias, stochastic and cosmological parameters, leading to a more diagonal cross-correlation matrix between these parameters \cite{Mergulhao:2021kip, Mergulhao:2023zso}.
On the other hand, the tracer split enhances the shot-noise contribution since $(\bar{n}^{t_i})^{-1}\geq (\bar{n}^{T})^{-1}$, which can deteriorate the signal-to-noise ratio. 

The counter-term for the MT case is
\begin{eqnarray}
    P_{\rm c.t.}^{t_it_j} &=& -\left[b^{t_i}_{\nabla^2\delta}(z)b^{t_j}_\d(z)+b^{t_j}_{\nabla^2\delta}(z)b^{t_i}_\d(z)\right]\frac{k^2}{k^2_{\mathrm{norm}}}\Plin(z) \; ,
\end{eqnarray}
and the stochastic is given by
\begin{equation}\label{eq:cross_stoc_def}
   P_{\rm stoch}^{t_it_j}(k,z) = \frac{1}{\sqrt{\bar{n}^{t_i}\bar{n}^{t_j}}}\left[\delta^K_{ij} + c^{t_it_j}_0(z) + c^{t_it_j}_2(z)\frac{k^2}{k^2_{\mathrm{norm}}} \right] \, ,
\end{equation}
where $\delta^K_{ij}$ is the Kronecker delta.  
It is often assumed that $P_{\rm stoch}^{t_it_j}(k,z) = 0$ for $j\neq i$, which implies that the stochastic fields $\epsilon^{t_i}$ and $\epsilon^{t_j}$  do not correlate, as small scale processes for the tracer $i$ are, in principle, independent of those for the tracer $j$. 
Noise suppression can substantially enhance the information extracted out of galaxy clustering \cite{Seljak:2009af}. 
Despite the cross-noise being relatively small compared to the diagonal part \cite{Hamaus:2010im}, as predicted by the halo model \cite{Cooray:2002dia}, it has been shown that exclusion effects, satellite galaxies \cite{Smith:2006ne} and also nonlinearities can enhance correlations on large scales \cite{Baldauf:2013hka}. 
While the cross-stochastic terms prevent the information gain in linear regime \cite{Gil-Marin:2010ezc} (see also \cite{Bernstein:2011ju}),
\cite{Mergulhao:2021kip,Mergulhao:2023zso} have found that the inclusion of those terms do not sensibly affect the constraints on cosmological parameters {\it when considering non-linear theory}. As we see later, this is a big difference compared to the linear MT results. We discuss the impact of including the cross-stochastic term later in this work. 

In redshift space, the power spectrum depends on the angle $\mu = \frac{\hat{z} \cdot \vec{k}}{k}$ measured relative to the line-of-sight direction $\hat{z}$ 
\ba
P^{t_i{t_j}}(k,\mu, z) &= Z^{t_i}_1(k, z)Z^{t_j}_1(k, z)
\Plin(k, z) +  3Z_1^{t_i}(\vk, z)\Plin(k, z)\int_{\vq}Z_3^{t_j}(\vq,-\vq,\vk, z)\Plin(q, z)  \vs
&+ 2\int_{\vq}Z^{t_i}_2(\vq,\vk-\vq, z)Z^{t_j}_2(\vq,\vk-\vq, z)
\Plin(z,|\vk-\vq|, z)
\Plin(q, z)  \vs
&+  3Z_1^{t_j}(\vk, z)\Plin(k, z)\int_{\vq}Z_3^{t_i}(\vq,-\vq,\vk, z)\Plin(q, z) 
+ P_{\text{c.t.}}^{{t_i}{t_j}}(k,\mu, z)
+ P_{\varepsilon^{t_i}\varepsilon^{t_j}}(k,\mu, z)\,,
\ea
where we follow the notation used in \cite{Mergulhao:2023zso}. For the complete form of the $Z$ kernels, including their dependence on the bias parameters, see \cite{Perko:2016puo,Chudaykin:2020aoj}. 
The counter-term and stochastic power spectrum are given by \cite{Mergulhao:2023zso}\footnote{Notice we do not include $c^{t_j}_{{\rm ct},26}$ as in \cite{Mergulhao:2023zso}, since it is completely degenerate with the other terms. We acknowledge Oliver Philcox for pointing this out.}
\ba 
P_{\rm c.t.}^{{t_i}{t_j}}(k,\mu, z) &=  \frac{k^2}{k_{\rm norm}^2}\Plin(k, z)
\left[Z^{t_i}_1 (k, z)\left(c^{t_j}_{{\rm ct},20}(z) + c^{t_j}_{{\rm ct},22}(z)\mu^2 + c^{t_j}_{{\rm ct},24}(z)\mu^4  \right.\right.\\
& \left.\left. + c^{t_j}_{{\rm ct},44}(z)\frac{k^2}{k_{\rm norm}^2}\mu^4 + c^{t_j}_{{\rm ct},46}(z)\frac{k^2}{k_{\rm norm}^2}\mu^6\right) + {t_i}\leftrightarrow {t_j} \right]\,, 
\vs
P_{\varepsilon^{t_i}\varepsilon^{t_j}}(k,\mu,z) &= \frac{1}{\sqrt{\bar{n}_{t_i}\bar{n}}_{t_j}}\left[c^{{t_i}{t_j}}_{\rm st, 00}(z) + c^{{t_i}{t_j}}_{\rm st, 20}(z)\frac{k^2}{k_{\rm norm}^2} + c^{{t_i}{t_j}}_{\rm st, 22}(z)\frac{k^2}{k_{\rm norm}^2}f(z)\mu^2 \right] \label{eq:PepsMT} \,.
\ea
Following \cite{Ivanov2019,Nishimichi:2020tvu,Mergulhao:2023zso}, we include higher-order terms in $k^2$ in the counter-term as a proxy for higher-order contributions to the Fingers-of-God (FoG) effect. 
We can then expand the power spectrum into multipoles
\be
P^{{t_i}{t_j}}(k,\mu, z) =\sum_{\ell \; {\rm even}} {\cal L}_\ell(\mu) P^{{t_i}{t_j}}_{\ell}(k, z)\,,
\ee
such that each multipole is given by the projection into Legendre polynomial
\ba
P^{{t_i}{t_j}}_{\ell}(k, z) \equiv \frac{2\ell+1}{2}\int_{-1}^{1}d\mu \, {\cal L}_\ell(\mu) 
P^{{t_i}{t_j}} (k,\mu, z)\,.
\ea
We use a modified version of \texttt{CLASS-PT} \cite{Chudaykin:2020aoj}, built on top of \texttt{CLASS} \cite{blas2011cosmic}, to compute these spectra, along with the inbuilt IR-resummation (see \cite{Senatore:2014via, Blas:2016sfa} for details).

\subsection{Fisher analysis}
\label{sec:fisher}

In real space, the Fisher information matrix for a single tracer $T$ is given by
\be
F_{\theta_a\theta_b}^{\rm ST} = \sum_{\alpha,\beta} \frac{\partial}{\partial \theta_a}P^{TT}(k_\alpha)  \, \Cov^{-1}_{TT,TT}(k_\alpha, k_\beta) \,\frac{\partial}{\partial \theta_b} P^{TT}(k_\beta) \,,
\ee
where $\boldsymbol{ \theta}$ is the vector of parameters, and the sum runs over the different discrete wavelength modes $k_\alpha$ and $k_\beta$. 
The (marginalized) error in the parameter $\theta_a$ is given by the square root of the inverse of the Fisher matrix
\be
\sigma_{\theta_a} = \sqrt{\left(F^{-1}\right)_{aa}}\,.
\ee
Moreover, we assume a diagonal Gaussian covariance 
\be
\Cov_{TT,TT}(k_\alpha, k_\beta) = \d_{\alpha, \beta}^K\frac{2}{m_{k_\alpha}}\left[P^{TT}(k_\alpha)\right]^2 \,,
\ee
where the shot noise is included in $P^{TT}$ as in \refeq{Pst}. Also, $m_{k_\alpha} = V k_\alpha^2 \Delta k / (2\pi^2) $ is the number of modes, where $V$ is the survey volume and $\Delta k$ is the $k$ bin width of the $k$-bin considered. For extensions of the Gaussian covariance, see \cite{Wadekar:2019rdu}. 

The extension of the Fisher information matrix for a MT set with $N_{\rm Tracers}$ tracers $t_1, \dots, t_{N_{\rm Tracers}}$ is given by \cite{Mergulhao:2021kip} (see also \cite{Abramo:2015iga})
\be \label{eq:FisherMT}
F_{\theta_a\theta_b}^{\rm MT} = \sum^{N_{\rm Tracers}}_{\substack{t_i,t_j,t_k,t_m, \\ {\rm with} \\ i\leq j; k \leq m}}\sum_{\alpha,\beta} \frac{\partial}{\partial \theta_a}P^{t_it_j}(k_\alpha)  \, \Cov^{-1}_{t_it_j,t_kt_m}(k_\alpha, k_\beta) \,\frac{\partial}{\partial \theta_b} P^{t_kt_m}(k_\beta)\,,
\ee
with the auto and cross-spectra calculated as \refeq{generalP} and the covariance
\ba
& \Cov_{t_it_j,t_kt_m}(k_\alpha, k_\beta)  = \d_{\alpha, \beta}^K \frac{1}{m_{k_\alpha}}\left[ P^{t_it_k}(k_\alpha) P^{t_jt_m}(k_\alpha) + P^{t_i t_m}(k_\alpha) P^{t_jt_k}(k_\alpha) \right] \,.
\ea
We also assume that all the tracers overlap within a single redshift bin{, taken for simplicity to be at redshift z=0}. Once more, we consider a Gaussian covariance for MT, since we restrict our analysis to relatively large scales $k \leq 0.15 \,h/$Mpc where the Gaussian covariance shows good agreement with the covariance measured from simulations \cite{Blot_2019}. {Moreover, since we focus on the gains of MT over ST and not in the overall gain by moving towards smaller scales, we work on the assumption that non-Gaussian corrections to the covariance affect the signal-to-noise of ST and MT similarly. We leave the investigation of the extend to which this assumption is valid and  the impact of the non-Gaussian covariance on MT, in the same lines as \cite{Wadekar:2019rdu}, for a future project. Furthermore, we discuss the stability of the Fisher derivatives in \refapp{derivatives}.}

The Gaussian covariance in redshift space can be written using the 3j Wigner symbols as\footnote{Some factors have been corrected relative to the derivation of \cite{Mergulhao:2023zso}.}
	\ba \label{eq:Covrsd}
	\Cov_{t_it_j,t_kt_m}^{\ell \ell'}(k_\alpha, k_\beta)  =&\d_{\alpha, \beta}^K
	\frac{\left( 2 \ell +1 \right) \left( 2 \ell' +1 \right)}{m_{k_\alpha}} 
	\sum_{\ell_1 \ell_2 \ell_3}  \left( 2 \ell_3 +1 \right) 
	\left( \begin{array}{ccc} \ell_1 & \ell_2 & \ell_3 \\ 0 & 0 & 0 \end{array} \right)^2 
	\left( \begin{array}{ccc} \ell & \ell' & \ell_3 \\ 0 & 0 & 0 \end{array} \right)^2
	\nonumber \\
	&
	\qquad
	\left[ 
	\left( -1 \right)^{\ell'} P_{\ell_1}^{t_it_k} \left( k_\alpha \right) P_{\ell_2}^{t_jt_m} \left( k_\alpha \right) 
	+  P_{\ell_1}^{t_it_m} \left( k_\alpha \right)   P_{\ell_2}^{t_jt_k} \left( k_\alpha \right) 
	\right]\,.
	\ea
The Fisher is then given by \refeq{FisherMT} summing over the monopole, quadrupole and hexadecapole contributions. Notice that the triangle inequality of the 3j Wigner symbols bounds $\ell_3$ to $|l_1 - l_2| \le l_3 \le |l_1 + l_2|$. In our analysis we consider only the first three even multipoles of the tracer power spectrum, so that the sums above are all effectively bounded.
{
We do not include light cone effects and the Alcock-Paczynski distortion in the analysis, since we work at a constant redshift of $z=0$.} 

\subsection{Fisher Setup} \label{sec:setup}

For the Fisher analysis, we adopt a Planck 18 cosmology \cite{Planck:2018vyg}, keeping $h, \omega_{\rm cdm}$ and $A_s$ free together with the bias parameters and without adding any priors.
A realistic tracer split accounts for observed features such as the color, sample mass, star formation rate or local overdensity, which, in terms of the bias expansion, result in different bias values for the MT subsamples. 
Here we bypass the split based on a specific tracer feature and instead directly consider tracers with different bias parameters. For the single-tracer case, we use for the bias parameters at $z=0$
\ba
b_1^{\rm ST} &= 1.16\, , \quad 
b_{\delta^2}^{\rm ST}  = b_{\mathcal{G}_2}^{\rm ST} = b_{\Gamma_3}^{\rm ST} = c^{\rm ST}_{{\rm ct},20}= 0.1 \,, \label{eq:fiducialvalues}
\ea
and all other stochastic and counter-terms fixed at 0.
We chose this fiducial parameter point such that the second and third-order operators are non-negligible but still small relative to the linear bias.
We find that the Fisher error shows some dependence on the choice of fiducial bias parameters, as we discuss in \refapp{fiducial}. 
Unless stated otherwise, the largest mode considered in the analysis is $k_{\rm max} = 0.15 \,h/$Mpc, with a tracer number density of $\bar{n}^T = 10^{-3} \,h^3 {\rm Mpc}^{-3}$, a survey volume $V = 11.9 \,{\rm Gpc}^3 h^{-3}$, comparable to DESI (see \reftab{survey}) and $z=0$. We explore other volume and redshifts in \refsec{forecasts}. 
We also include the cross-stochastic contribution \refeq{cross_stoc_def} between two different tracers.
For the multi-tracer with two tracers, we define their bias parameters relative to the single tracer value as
\be \label{eq:deltab2}
b_{\O}^{t_1} = b_{\O}^{\rm ST} \mp \frac{\Delta b}{2}  \quad \textrm{and} \quad b_{\O}^{t_2} =  b_{\O}^{\rm ST} \pm  \frac{\Delta b}{2} \,,
\ee
and we display results for different $\Delta b$ values. 
Notice that it is also possible to consider a split in the counter-terms, associated in this case to different Lagrangian radius of the two tracers. This could lead to different perturbative behaviors in the expansion in terms of the higher-derivative operators, as we discuss in \refsec{fog}. 
Furthermore, unless stated otherwise, we consider a balanced split in the number density, such that
\be
\bar{n}^T = \sum_{t_i} \bar{n}^{t_i}\,
\ee
and 
\be
\bar{n}^{t_1} = \bar{n}^{t_2} = \dots = \bar{n}^{t_N}\,.
\ee

\section{The bias split and MT information} \label{sec:biassplit}
In this section, we examine how MT behaves under a direct split in the bias parameters.
In \refsec{biasdiff}, we discuss the dependence of the MT results on the difference between the final bias values of the subsamples. In \refsec{toymodel}, we explore how likely it is to find subsamples with different inear bias coefficients. Finally, in \refsec{info}, we analyze the MT information content and the importance of the cross-stochastic term.

\subsection{The dependence on the bias difference} \label{sec:biasdiff}

\begin{figure}[h]
	\centering
	\includegraphics[width = 0.8\textwidth]{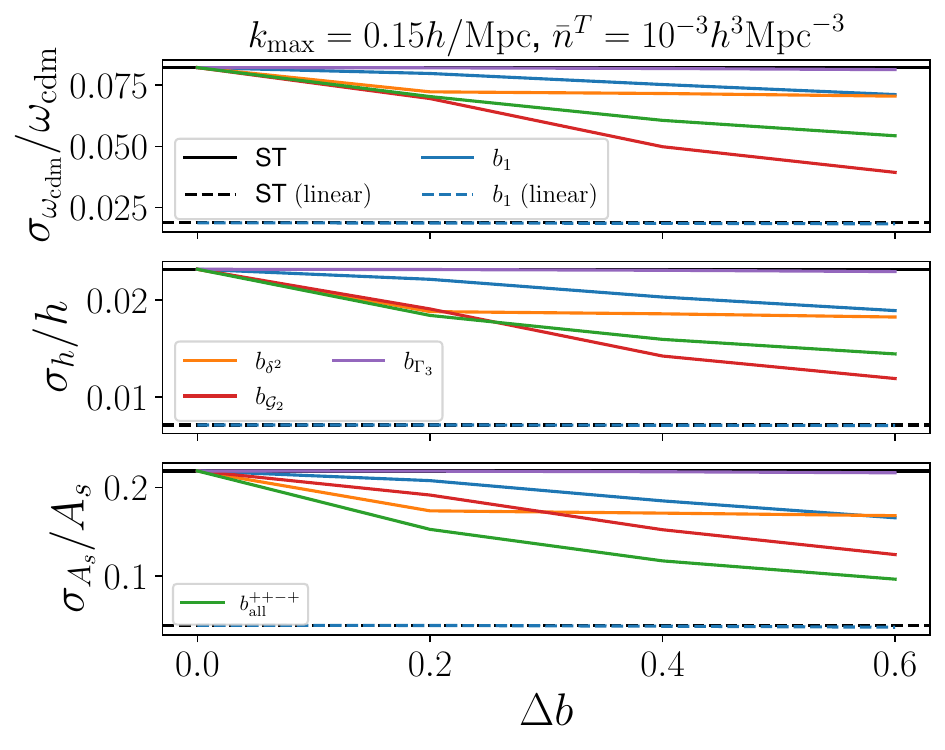}
	\caption{ Relative errors in $\omega_{\rm cdm}$, $h$ and  $A_s$ as a function of the difference in each bias parameter, $\Delta b$.  The solid (dashed) black line represents the single-tracer EFT (linear) results, while the dashed blue line corresponds to a multi-tracer split in $b_1$ in linear theory. The $b_{\rm all}^{++-+}$ curve indicates a simultaneous split in {\it all} four bias parameters, with the $\pm$ signs denoting the sign of $b_1$, $b_{\d^2}$, $b_{\G_2}$, $b_{\Gamma_3}$ for the $t_2$ sample, as described in \refeq{deltab2}. We consider $k_{\rm max} = 0.15 \,h/$Mpc (see \reffig{deltab_all} for $k_{\rm max} = 0.05 \,h/$Mpc) and assume a tracer number density $\bar{n}^T = 10^{-3} \,h^3 {\rm Mpc}^{-3}$. }
	\label{fig:deltab}
\end{figure}

We start by discussing the optimal multi-tracer split to maximize the information extracted from a sample. 
A key advantage of working within a Fisher framework is that we can directly assess the impact on the Fisher information by smoothly varying the $\Delta b$ parameters for every bias parameter separately.
\reffig{deltab} displays the relative errors in $\omega_{\rm cdm}$, $h$ and $A_s$ for both ST (black lines) and MT as a function of the bias difference $\Delta b$ [see \refeq{deltab2}]. 
The solid lines represent the results that include non-linear theory.
The $b_1$, $b_{\d^2}$, $b_{\G_2}$ and $b_{\Gamma_3}$ lines correspond to a tracer split in different biases while keeping all the other parameters fixed at the ST values from \refeq{fiducialvalues}. Notably, in the limit $\Delta b \to 0$ the results always match the ST case (black lines), indicating that a tracer split that does not introduce a bias difference results in no relative information gain. 
\reffig{deltab} shows that a tracer split in $\Delta b_{1}, \Delta b_{\d^2}$ and  $\Delta b_{\G_2}$ significantly improves the relative error bars for the cosmological parameters compared to the ST analysis, with $\Delta b_{\G_2}$ yielding the greatest information gain. However, the gains are minor for a split in the third-order operator $b_{\Gamma_3}$.\footnote{The comparison between the relative gains for each of the bias parameters as a function of $\Delta b$ is complicated, since one can always change base of the bias expansion by taking linear combinations of the operators $\mathcal{O}$. Then, also the biases change correspondingly.
Instead, a more meaningful question is how hard it is to obtain a $\Delta b$ for a given operator in a given basis, which we address in \refsec{toymodel}.}
Additionally, we present extra plots in \refapp{extraplots}, where we {analyze the dependence on $k_{\rm max}$. In the large-scale regime, we observe that the error bars increase significantly. Also, a split in $b_1$ outperforms a split in $b_{\d^2}$ or $b_{\G_2}$ since the linear term becomes dominant}. This suggests that, as smaller scales are included in the analysis, the importance of splitting in non-linear operators grows. We also notice some dependence of the results on the fiducial point chosen for the bias parameters, as discussed in \refapp{fiducial}. The qualitative results, however, remain unchanged. 

For comparison, the dashed lines in \reffig{deltab} represent the results obtained using linear theory. 
Naturally, the error bars are smaller in that case, as fewer parameters need to be marginalized over.  However, the absence of EFT corrections may lead to biased cosmological parameters, given that we are considering relatively nonlinear scales. 
The blue dashed line indicates a split in $b_1$ within linear theory, where we see a very mild dependence on $\Delta b_1$. 
As we discuss in \refsec{info}, MT provides improved results under a linear split only when the cross-stochastic term is neglected or for very high values for $\bar{n}^T$. 
We find that a split in $\Delta b_1$ yields significantly {smaller $\sigma_{\rm MT}/\sigma_{\rm ST}$} when non-linear corrections are included {(both for ST and MT)}, even when accounting for cross-stochasticity.
This improvement is likely to happen due to extra degeneracy breaks in the non-linear operators (see \refsec{info}), e.g.~in terms proportional to $b_{1}b_{O}$.

\begin{figure}[h]
	\centering
	\includegraphics[width = 0.99\textwidth]{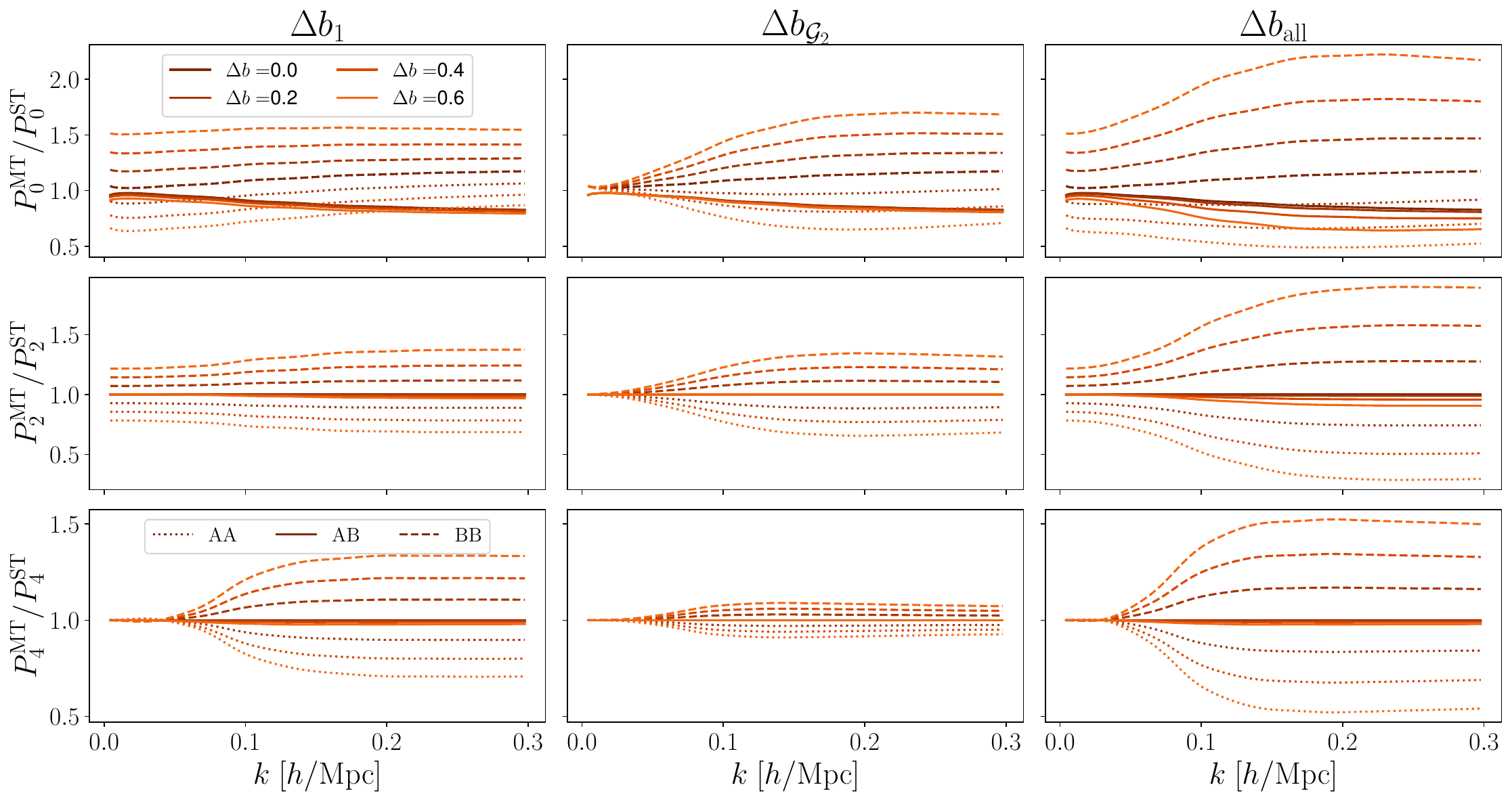}
	\caption{ Monopole, quadrupole and hexadecapole normalized by the ST spectra for a bias split in {$b_1$ (left), $b_{\G_2}$ (middle) and $b_{\rm all}$ (right)}. Different colors correspond to various $\Delta b$ value. Dotted, solid and dashed lines correspond to the auto AA, cross AB and auto BB spectra, respectively.  }
	\label{fig:multipoles}
\end{figure}

{So far we have only discussed MT splits in orthogonal directions in the bias parameter space, namely changing one bias parameter at a time. However, a realistic tracer split based on a physical motivated feature is going to generate a split in {\it all} the linear and non-linear bias parameters. To account for this scenario, we introduced a homogeneous tracer split in {\it all bias parameters simultaneously}, denoted in \reffig{deltab} as $b_{\rm all}^{++-+}$.}
The four $\pm$ signs in the superscript represent the signs of $b_1$, $b_{\d^2}$, $b_{\G_2}$, $b_{\Gamma_3}$ for the $t_2$ sample, as indicated by \refeq{deltab2}.
Interestingly, this split does not necessarily lead to better results compared, for example, to a split in $b_{\G_2}$, particularly for $\omega_{\rm cdm}$ and $h$. 
This suggests that the Fisher information may be more sensitive to non-trivial parameter combinations rather than the simple variation of all bias coefficients in a single direction. 
 In \reffig{deltab_all} of \refapp{extraplots}, we also present results for ${\Delta} b_{\rm all}^{+\pm\pm\pm}$, with different sign combinations. 
 In general, we find that splitting all four bias parameters together consistently performs better than a simple split in the linear bias $b_1$. From here on, we show results as a function of ${\Delta} b_{\rm all}^{++-+}$ (hereafter referred to simply as $ {\Delta} b_{\rm all}$), where the sign of $b_{\G_2}$ is flipped, as a proxy of a generic tracer split.\footnote{{The optimal combination of bias parameters that maximizes the MT results is a parameter-dependent problem. Note that the sign convention for the bias (e.g., having a minus sign for $b_{\G_2}$) is also arbitrary, since linear combinations of operators can be taken to form a new basis. Also note that the definition of the operators depend on the chosen cutoff scale for the integrals, as well as on the renormalization scheme \cite{Rubira:2023vzw}. The optimal combination of bias parameters to maximize MT performance for a given parameter is to be explored in future work and the choice of the minus sign for $b_{\mathcal{G}_2}$ is related to the fact that it performs on average for $\omega_{\rm cdm}$ and $h$, and reasonably well for $A_s$ (see \reffig{deltab_all}).}
 }
 We display in \reffig{multipoles} the monopole, quadrupole and hexadecapole spectra for the MT scenario relative to the ST case. The left panel shows the effects of a split in $b_1${, for the middle panel we consider a split in $b_{\G_2}$} and the right panel illustrates the case of a split in $b_{\rm all}$, where all bias parameters are varied simultaneously by the same $\Delta b$. Notice that the {MT monopole is not exactly the same as the ST monopole for $\Delta b = 0$} due to the $\frac{1}{2}$ factor of difference in the number density, which enters in the stochastic part, {cf.~\refeq{PepsMT}}. {While $b_{1}$ is responsible for a major vertical shift in the spectra, the split in $b_{\G_2}$ leads to a scale-dependent feature that becomes important when analyzing the broadband spectra. It is also understandable that the impact of $b_{\G_2}$ is larger than a split in $b_{\d^2}$ and $b_{\Gamma_3}$. While the latter only starts to contribute at third order in perturbation theory, $b_{\d^2}$ has its third-order contribution, i.e.~its contraction with $\delta^{(1)}$, removed by renormalization (see Eq.~4.11 of \cite{Assassi2014}). Therefore, $b_{\G_2}$ is the only non-linear bias operator that contributes {\it both} at second and third order.}

We summarize the results in \reftab{improv}, where we fixed $\Delta b = 0.6$. We find that a split in $b_1$ using linear only results in a $5\%$ improvement in $A_s$ with barely any gain for the other two cosmological parameters. 
On the other hand, considering a split in $b_1$ and using non-linear theory leads to improvements of $13\%$ in $\omega_{\rm cdm}$, $18\%$ in $h$ and $24\%$ in $A_s$. 
When split in the full set of bias parameters simultaneously, the improvements reach $34\%$ in $\omega_{\rm cdm}$, $38\%$ in $h$ and $56\%$ in $A_s$. Those values are comparable to the findings of \cite{Mergulhao:2021kip,Mergulhao:2023zso}.
Thus, we find that selecting samples with significantly different bias parameters can considerably enhance cosmological constraints, sometimes reducing the error bars by a factor two or more. 
\begin{table}[!htb]
	\begin{minipage}{1\linewidth}
		\centering
		\begin{tabular}{|c||c|c|c|} \hline 
			 & $b_1$ (linear) & $b_1$ (EFT) & $b_{\rm all}$ (EFT) \\ \hline \hline
			$\sigma_{\omega_{\rm cdm}}^{\rm MT}/\sigma_{\omega_{\rm cdm}}^{\rm ST}$   & 0.98& 0.87& 0.66   \\ \hline
			$\sigma_{h}^{\rm MT}/\sigma_{h}^{\rm ST}$   &  0.99&0.82&0.62 \\ \hline 
			$\sigma_{A_s}^{\rm MT}/\sigma_{A_s}^{\rm ST}$   &  0.95&0.76& 0.44  \\ \hline 
		\end{tabular}
	\end{minipage}%
	\caption{MT improvement relative to the ST result for $\Delta b = 0.6$, shown for a split in the linear bias (both linear and non-linear modeling) and a split in $b_{\rm all}$. We assume $\bar{n}^T = 10^{-3} \,h^3 {\rm Mpc}^{-3}$. }
	\label{tab:improv}
\end{table}

\subsection{The bias split using the halo mass} \label{sec:toymodel}

In this section, we discuss how likely it is to find a difference $\Delta b$ in a galaxy sample, putting the results of the previous section into perspective. We anticipate that, in a vanilla scenario, it is very difficult to find a large difference in the (linear and non-linear) bias, but assembly bias can substantially enhance this difference to potentially reach $\Delta b \sim 1$.
To do so, we consider a halo sample following Tinker's mass function \cite{Tinker:2008ff} and linear bias \cite{Tinker:2010my}. We restrict our analysis to halos of masses in between $10^{13} M_{\odot}/h$ and $ 10^{15} M_{\odot}/h$, which typically lead to larger bias values.
The values of $b_{\d^2}$ are determined using its functional fit from separate Universe simulations \cite{Lazeyras:2015lgp} for halos 
\be \label{eq:bd2}
b_{\d^2}(b_1) = 0.206 - 1.071b_1 + 0.464b_1^2 + 0.004b_1^3\,.
\ee

\begin{figure}[h]
	\centering
	\includegraphics[width = 0.49\textwidth]{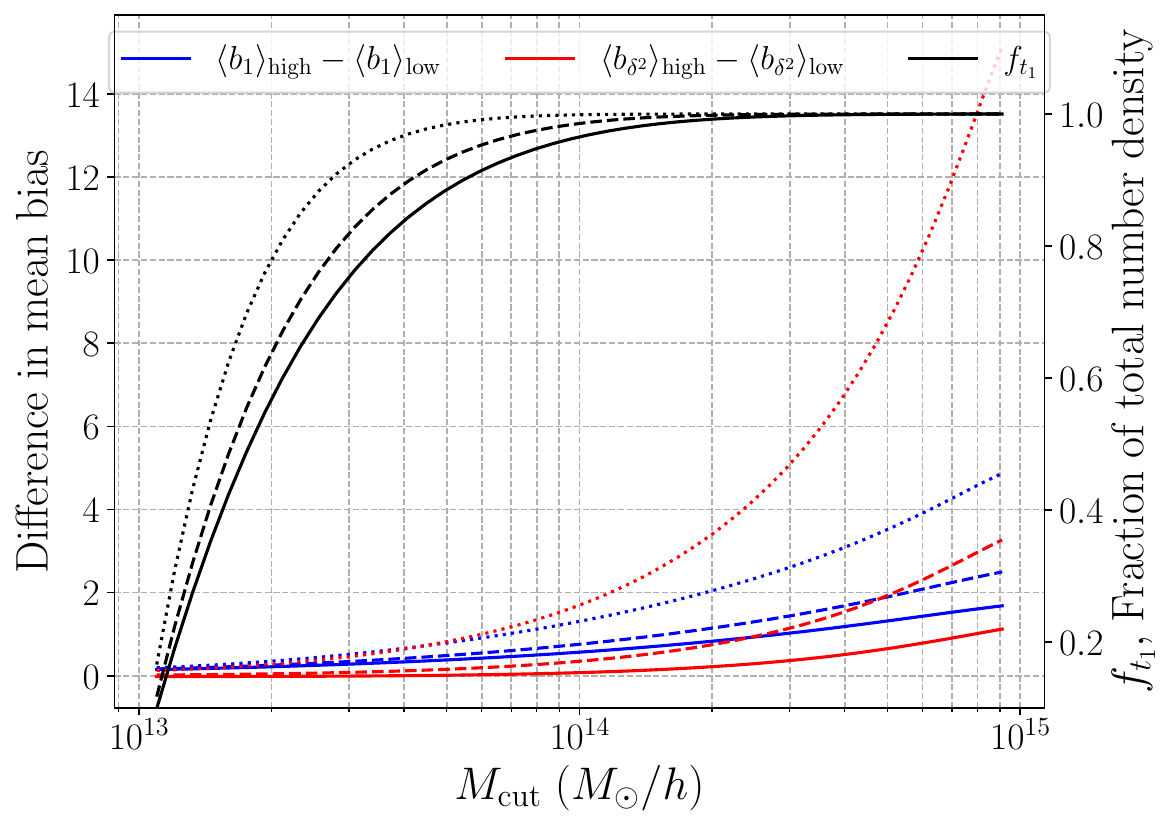}
	\includegraphics[width = 0.47\textwidth]{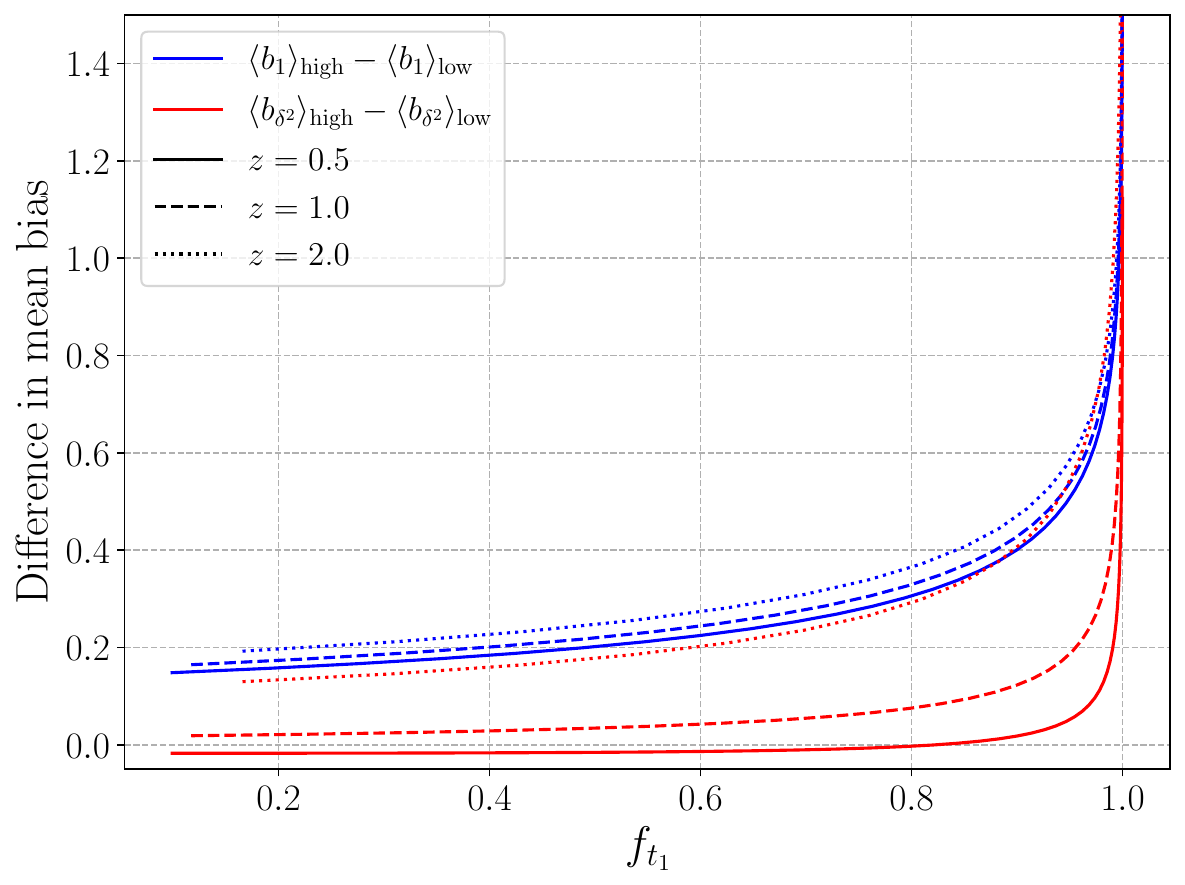}
	\caption{ In the left panel, the difference in the mean values of $b_1$ (blue) and $b_{\d^2}$ (red) between the lower and higher mass samples defined by a cut $M_{\rm cut}$. Different redshifts are represented by different line styles (solid, dashed and dotted lines). We show $f_{t_1}$, the density fraction of the lower mass tracer $t_1$ [see \refeq{frac}], as black lines, with its values indicated on the right y-axis. In the right panel, the same differences are plotted as a function of  $f_{t_1}$.}
	\label{fig:biassplit}
\end{figure}

We consider a threshold $M_{\rm cut}$ to divide the sample into two subsets with different $b_1$ and $b_{\d^2}$ distributions.
The quantities $\langle b_1 \rangle_{\rm low}$ and $\langle b_1 \rangle_{\rm high}$ represent the mean value of $b_1$ in the lower and higher-mass subsets, respectively, where the expectation value is weighted by the Tinker's mass function (using the values for $\Delta = 200$ of \cite{Tinker:2008ff}). Similarly, $\langle b_{\d^2} \rangle_{\rm low}$ and $\langle b_{\d^2} \rangle_{\rm high}$ represent the mean quadratic bias for each subset.
We show the difference between the higher-mass and lower-mass biases as a function of $M_{\rm cut}$ for different redshifts (solid, dashed, dotted lines) in the left panel of \reffig{biassplit}. 
We also display as black lines the fraction of the tracer $t_1$ number density relative to the total number density 
\be \label{eq:frac}
f_{t_1} = \frac{\bar{n}^{t_1} }{\bar{n}^{T}} \,.
\ee
We plot in the right panel of \reffig{biassplit} the bias difference as a function of $f_{t_1}$. 
This exercise, focused on halos split by their mass, shows that it is relatively hard to obtain a large difference in $b_1$ and $b_{\d^2}$ between the samples, unless one considers large values of $f_{t_1}$ or high redshift. As we discuss later in \refsec{shot}, the MT information gains relative to ST are present as long as $0.1\lesssim f_{t_1} \lesssim 0.9$.
Although this range allows for relatively unbalanced samples and offers more freedom to find subtracers with different bias, our vanilla analysis for halos indicates that it is difficult to have large non-linear bias differences at low redshift.\footnote{When considering higher-$z$, the difference in $b_{\d^2}$ between the samples increases. Thus, high-redshift samples are an interesting target to search for very different non-linear coefficients, such as considered in \cite{Ebina:2024ojt}.} 
Finally, we comment on a possible split in $b_{\G_2}$ or $b_{\Gamma_3}$. Notice that a split in $b_{\G_2}$ was among the most effective for MT in \reffig{deltab}. When we restrict ourselves to the standard local-in-matter-density (LIMD) expression for halos \cite{chan2012gravity,Desjacques:2016bnm}
\be
b_{\G_2}(b_1) = -\frac{2}{7}(b_1-1) \,,
\quad
\textrm{and}
\quad
b_{\Gamma_3}(b_1) = -\frac{23}{42}(b_1-1)
\,,
\ee
this immediately implies a factor $2/7$ and $23/42$ suppression in the difference in $b_{\G_2}$ and $b_{\Gamma_3}$, respectively, relative to $b_1$.

The above results indicate a very pessimistic scenario for finding tracers with different non-linear bias parameters. However, we should keep in mind that these findings are restricted to halos that follow either the separate Universe fits of \refeq{bd2} or the LIMD relations.
In a more realistic scenario, several effects can enhance the difference between samples. First, assembly bias can be very relevant, particularly for $b_{\G_2}$; for instance, \cite{Lazeyras:2021dar} has shown that $b_{\G_2}$ strongly deviates from the LIMD relation, indicating the possibility of finding very strong dependence of $b_{\G_2}$ on other properties beyond the halo mass. 
{Thanks to assembly bias, finding a difference $\Delta b_{\G_2} \sim 1.0$ appears to be realistic, at least in the situation of samples with large $b_1$ (or correspondingly large halo masses), see Fig.~5 of \cite{Lazeyras:2021dar}.}

{We highlight that \cite{Lazeyras:2021dar} considered halo masses $\sim 10^{14}\Msunh$, larger e.g.~than target values for DESI \cite{Yuan:2023ezi}, and therefore with large $b_1$. The feasibility of identifying samples with very distinct non-linear bias for low-$b_1$ cases still requires further investigation.} Second, realistic galaxies samples may also strongly deviate from the relations that hold for halos \cite{Voivodic:2020bec}. A more comprehensive study on the feasibility of finding two samples with distinct nonlinear biases is beyond the scope of this work, and we leave it for a future project. {Finally, we note that assembly bias can also be used to break the degeneracy between $b_1$ and the non-Gaussian bias parameter $b_\phi$, improving the MT constraints in $f_{\rm NL}$ \cite{Fondi:2023egm,Barreira:2023rxn,Karagiannis:2023lsj}.} 

\subsection{The multi-tracer information content} \label{sec:info}

In this section, we investigate the MT information content.
A central aspect of MT analysis is that the dataset is rearranged so that, instead of considering a single spectrum, we have
\be
\frac{N_{\rm Tracers} \times (N_{\rm Tracers} +1 )}{2}\,,
\ee  
spectra, which includes both auto and cross-spectra between different tracers.
The trade-off is that the number of biases, counter-terms, and stochastic parameters increases. While in the ST case we have $N_{\rm bias}$, $N_{\rm ct}$ and $N_{\rm stoch}$, in the MT these grow to
\ba
N_{\rm Tracers} \times N_{\rm bias}& \quad \textrm{bias parameters} \,, \\
N_{\rm Tracers} \times N_{\rm ct}& \quad \textrm{counter-terms} \,, \\
\frac{N_{\rm Tracers} \times (N_{\rm Tracers} + 1)}{2} \times N_{\rm stoch}& \quad \textrm{stochastic parameters} \,.
\ea
Whether the expanded data array in MT compensates for the increased number of parameters is a fundamental question in assessing its relative information gain.

\begin{figure}[h]
	\centering
	\includegraphics[width = 0.85\textwidth]{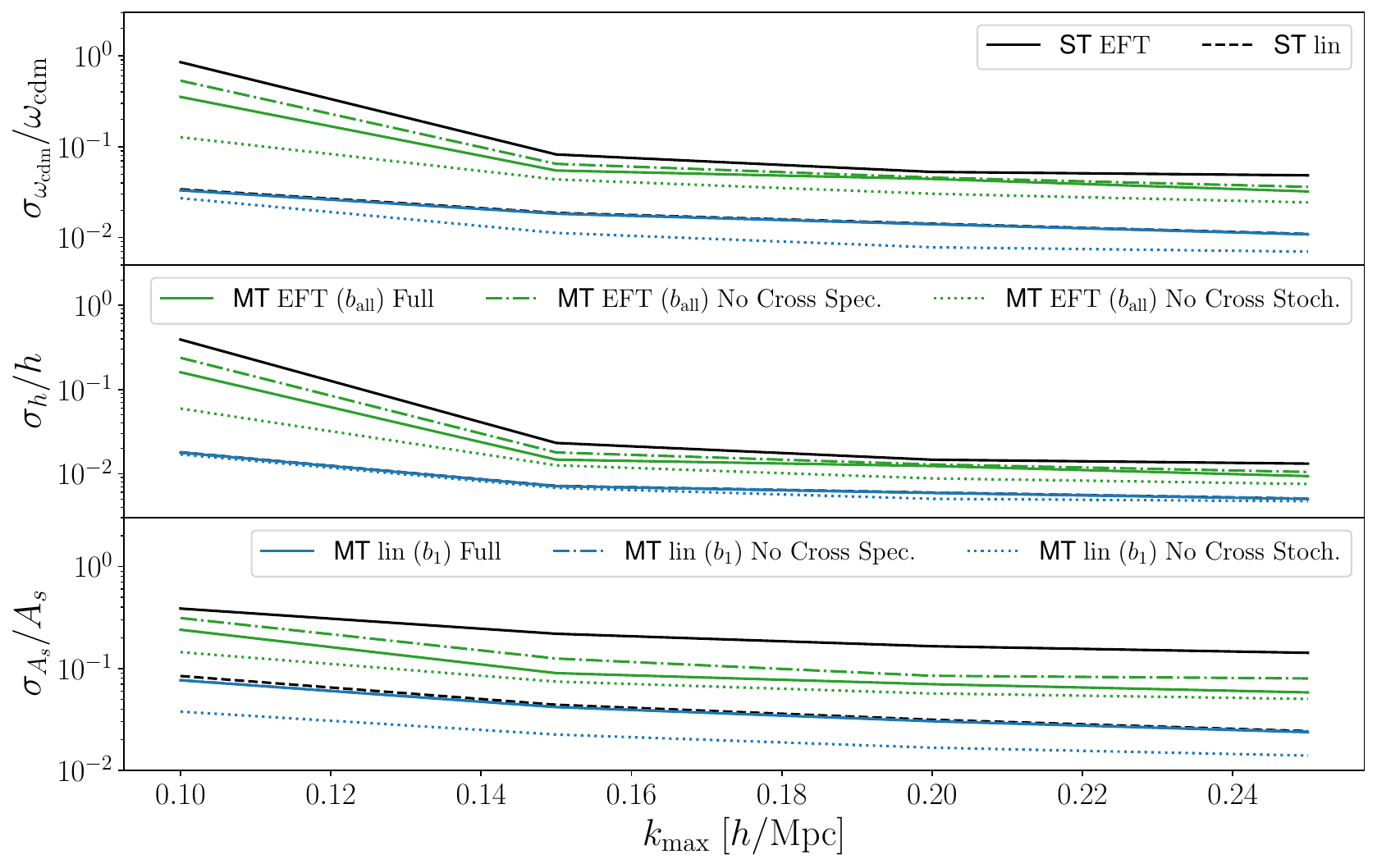}
	\caption{ Relative error in {$\omega_{\rm cdm}$, $h$ and }$A_s$ as a function of the maximum mode $k_{\rm max}$ included in the analysis. The solid lines represent the full MT analysis, which includes both the cross spectra and the cross stochastic terms. When the cross-stochastic parameters are removed (dotted lines), the error bars decrease, while completely removing the cross spectra (dash-dotted lines) leads to an increase in the error bars. {We fix either $\Delta b_1 = 0.6$ or $\Delta b_{\rm all} = 0.6$.} 
	}
	\label{fig:cross}
\end{figure}

We start by discussing the relevance of the cross spectra.
We show in \reffig{cross} the relative error in {$\omega_{\rm cdm}$, $h$ and} $A_s$ as a function of the maximum $k_{\rm max}$ mode considered in the analysis, fixing the tracer number density $\bar{n}^T = 10^{-3} \,h^3 {\rm Mpc}^{-3}$. 
We compare the MT and ST case in two different scenarios: one with a linear modeling and a split in $b_1$ and another with a nonlinear modeling and with a split in $b_{\rm all}$.
The solid lines describe the full analysis, including both the cross-spectra and the cross-stochastic terms. In contrast, the dashed-dotted lines correspond to a case where we removed the cross-spectra between the two tracers, keeping only the auto correlations in \refeq{FisherMT}. 
For the EFT case, we observe that removing the cross spectra leads to a loss of information, with the result approaching the ST case.  
However, the gains are not {\it completely} lost when removing the cross-spectra, indicating that MT still overcomes ST even when considering the auto spectra alone.  
For the linear case, we notice that the MT full-analysis line closely follows the ST case. Moreover, removing the cross-spectra results in no information loss, as the two solid and dashed-dotted lines completely overlap.

Another key question in MT analysis is whether cross-stochasticity in the cross-spectra can be neglected.\footnote{Dropping of cross-stochasticity is often justified by assuming that small-scale processes of the two tracers do not correlate. However, this assumption was shown to fail due to non-linear clustering and exclusion effects \cite{Baldauf:2013hka}. An alternative to completely removing this term is to introduce simulation-based priors on this cross-stochasticity \cite{Kokron:2021faa}. }
As discussed earlier, the linear case shown in \reffig{cross} suggests that MT provides almost no improvement over ST {\it when cross-stochasticity is included} in the cross spectra [\refeq{cross_stoc_def} for $i \neq j$].
For comparison, we display in dotted lines the same scenario but with the cross-stochastic contribution removed. In this case, a considerable information gain is observed. This indicates that the success of MT {\it in linear theory}  seems to strongly depend on the assumption of no cross-stochasticity between the tracers{, as already pointed out by other works \cite{Gil-Marin:2010ezc,Bernstein:2011ju}}.
{As we discuss later in this section, the removal of the cross-stochastic leads to strong breaking of degeneracies between parameters via the cross-spectra information, e.g., by fixing the $b_1^{t_1}b_1^{t_2}$ product. When the cross-stochastic term is included, this information gain is essentially lost.}
In the nonlinear case, we also observe an increase in information gain when removing the cross-stochastic parameters. However, different than the linear case, MT already shows a significant improvement over ST even when cross-stochasticity is included. {Extra information is therefore provided by the full $k$-dependence of the cross-spectra, allowing for a broader breaking of degeneracies among $b_{\O_\alpha}^{t_1}b_{\O_\beta}^{t_2}$, beyond the simple $b_1^{t_1}b_1^{t_2}$ that holds in the linear case.} This highlights a fundamental difference in the information budget of MT between linear and nonlinear theory, aligning with the findings of \cite{Mergulhao:2021kip, Mergulhao:2023zso}. 

\begin{figure}[h]
	\centering
	\includegraphics[width = 0.48\textwidth]{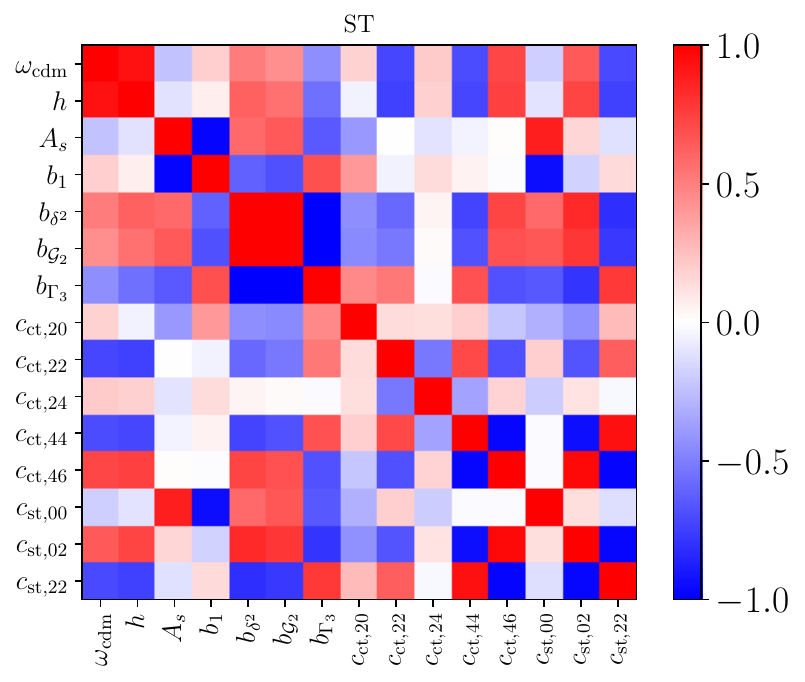} 
	\includegraphics[width = 0.48\textwidth]{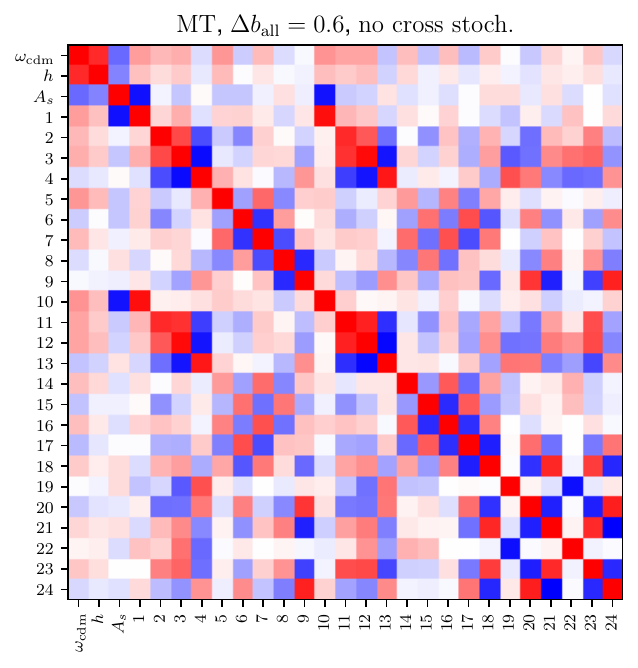}
    \includegraphics[width = 0.48\textwidth]{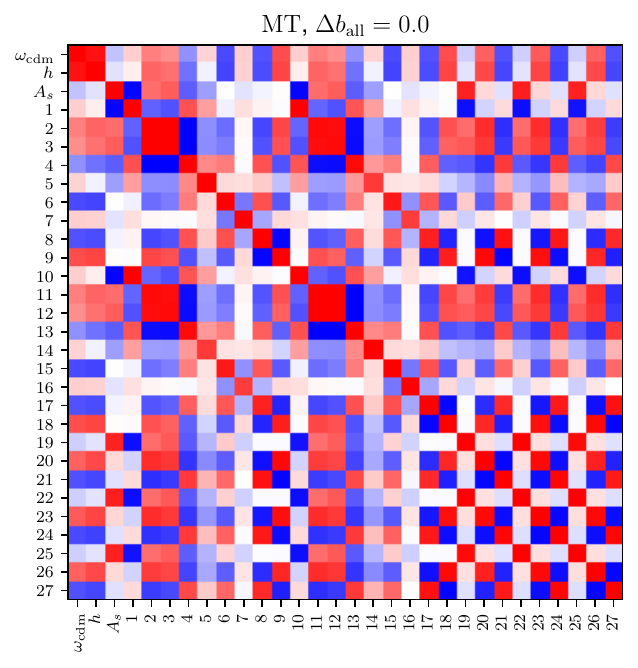}
	\includegraphics[width = 0.48\textwidth]{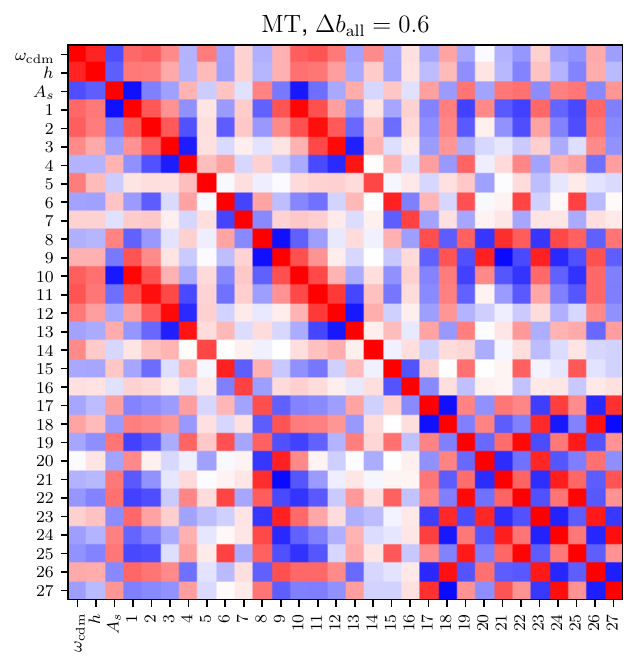}
	\caption{ {Correlation matrices between the parameters for the ST case (top left), and MT both without (top right, with $\Delta b_{\rm all} = 0.6$) and with (bottom left with $\Delta b_{\rm all} = 0.0$ and bottom right with $\Delta b_{\rm all} = 0.6$) the cross-stochastic operator. To avoid cluttering the MT matrices with labels, we denote the bias, counter-terms, and stochastic parameters by numbers, arranged in the same order as in the ST case (i.e., $b_{1}^{t_1}, b_{1}^{t_2}, b_{\d^2}^{t_1}, b_{\d^2}^{t_2}, \dots, c_{\rm st, 22}^{t_1t_1}, c_{\rm st, 22}^{t_1t_2}, c_{\rm st, 22}^{t_2t_2}$, with the stochastic terms in the end).}  }
	\label{fig:corrmatrix_new}
\end{figure}

We now investigate whether MT helps break degeneracies between free parameters. As pointed out in \cite{Mergulhao:2021kip,Mergulhao:2023zso}, the MT basis can be more diagonal, meaning that part of the information gain comes from choosing a parameter basis that better separate internal degeneracies. We show in \reffig{corrmatrix_new} the parameters (Pearson) correlation matrix 
\be
r_{ij} = \frac{(F^{-1})_{ij}}{\sqrt{(F^{-1})_{ii} (F^{-1})_{jj}}}\,,
\ee
 between the parameters $i$ and $j$. We present results for the ST case (top left) and for MT scenarios with a split in $b_{\rm all}$, both including (bottom) and excluding (top right) the cross stochastic term.  
Overall, we observe that cross correlations between cosmological and other parameters tend to be smaller in the MT case, particularly when cross-stochasticity is removed. {For the case including the cross-stochastic contribution, we show $\Delta b_{\rm all} = 0$ (equivalent to ST in information) in the bottom left and $\Delta b_{\rm all} = 0.6$ on the right. The latter is clearly more diagonal, strongly indicating that MT provides a consistent way to break correlations between bias and cosmological parameters.}

\begin{figure}[h]
	\centering
	\includegraphics[width = 0.9\textwidth]{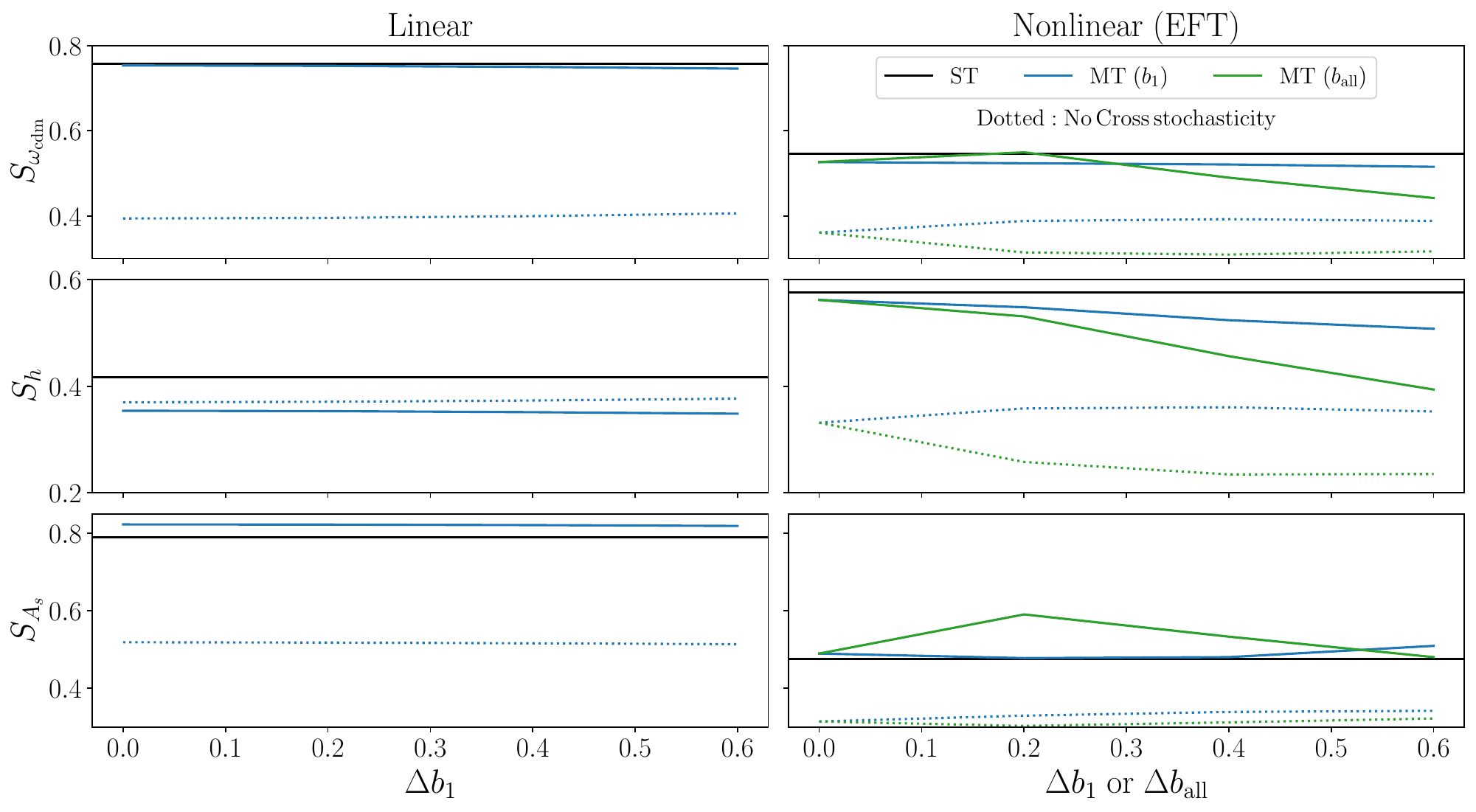}
	\caption{{Parameter correlation score defined in \refeq{score} as a function of $\Delta b$  for $\omega_{\rm cdm}$ (top), $h$ (middle) and $A_s$ (bottom). In the left panels, we consider linear theory, for which we find almost no dependence on $\Delta b$, while in the right panels we use the EFT non-linear modeling. The ST results are shown in black for comparison. The green line indicates a split in all bias parameters $\Delta b_{\rm all}$, while the blue line a simple split in $b_1$. Dotted lines show results with the cross-stochasticity removed. }  } 
	\label{fig:score}
\end{figure}

To quantify the degeneracy breaking, we define the cross correlation score for the parameter $\theta_i$ as 
\begin{equation} \label{eq:score}
S_{\theta_i}^2 = \frac{1}{N-1} \sum_{\substack{j=1 \\ j \neq i}}^{N} r_{ij}^2 \,,
\end{equation}
which represents the average squared value of the non-diagonal elements, with $N$ being the total number of parameters. Lower values indicate a more orthogonal basis for the parameter $i$. {We emphasize that a lower score does not necessarily implies better or worst constraining power, which is ultimately determined by the Fisher matrix $F$, but it can serve as a diagnostic to check whether degeneracy breaking can explain differences in the Fisher information.
We display this score for $\omega_{\rm cdm}, h$ and $A_s$ in the top, middle and bottom panels of \reffig{score}, showing results for three cases: linear modeling with a split in $b_1$ (left) and nonlinear modeling with a split in $b_1$ or $b_{\rm all}$ (right). The dotted lines are obtained discarding the cross-stochastic contribution. 
We see that removing the cross-stochasticity dramatically improves the degeneracy score in both linear and non-linear cases (with the exception of $S_h$ in the linear regime, which remains roughly unchanged). This indicates that in the absence of cross-stochasticity, the cross-spectra information substantially enhances the MT performance. As previously mentioned, adding motivated priors on cross-stochasticity may therefore help break degeneracies between parameters in MT.

In the linear case (left), we find almost no dependence on $\Delta b_1$ for all parameters, indicating that a large $b_1$ difference between tracers induces almost no degeneracy breaking. On the other hand, removing the cross-stochastic parameter significantly decreases the degeneracy score compared to ST for $A_s$ and $\omega_{\rm cdm}$, suggesting that the cross-spectra information in this case serves to break degeneracies and can explain the MT gains. 

In the non-linear modeling (right), removing the cross-stochasticity leads to major gains in degeneracy breaking, in agreement with the results presented in \reffig{corrmatrix_new}. Also, in opposition to the linear scenario, we find a strong dependence on $\Delta b$, particularly when considering a split $\Delta b_{\rm all}$. For $\omega_{\rm cdm}$ and $h$, the degeneracy score is substantially reduced as the bias difference increases, suggesting that this is a key factor in the improved MT constraints relative to ST. The decreasing degeneracy observed for MT indicates that a large bias difference leads to a more orthogonal basis for the cosmological parameters. Interestingly, the score for $A_s$ initially increases and then decreases again, despite decreasing error bars (see \reffig{deltab}), indicating that the improvement in $A_s$ may be caused by other effects, such as cosmic variance cancellation. 
}

\section{More tracers and the shot-noise dependence}
\label{sec:results}

	In this section we extend the results of the former section to consider more than two tracers in \refsec{moretracers} and different tracer number densities in \refsec{shot}.

\subsection{More than two tracers} \label{sec:moretracers}

A natural question is why not to further consider  splittings into more $N_{\rm Tracers}$ instead of just two tracers. There is, however, a trade-off between splitting into more tracers and the enhancement of the shot-noise contribution which eventually overtakes the total signal, since $\bar{n}^{t_i} = \bar{n}^{T}/N_{\rm Tracers}$, when considering all the subtracers with the same number density.
In this work, we explore for the first time the possibility of a split into more than two tracers when including non-linear parameters in the galaxy clustering modeling.

\begin{figure}[h]
	\centering
	\includegraphics[width = 0.7\textwidth]{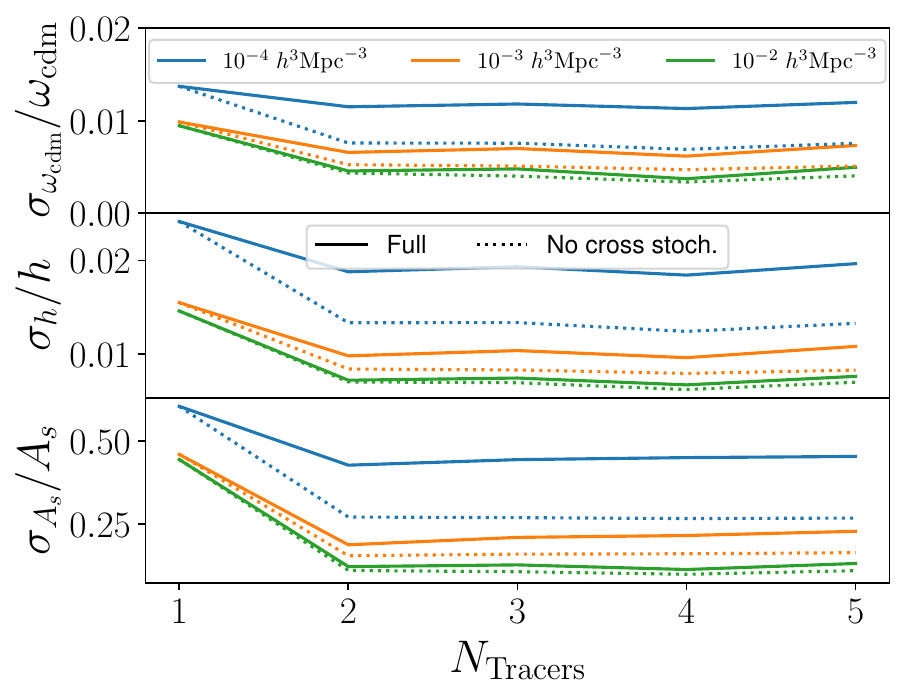}
	\caption{  Relative errors in $\omega_{\rm cdm}$, $h$ and  $A_s$ as a function of the number of tracers $N_{\rm Tracers}$ and for different $\bar{n}^T$ values. We fix $\Delta b_{\rm all} = 0.6$. For the solid lines, we include the cross-stochastic terms and we exclude them for the dashed lines. }
	\label{fig:ntracers}
\end{figure}

We display as solid lines in \reffig{ntracers} the evolution of the Fisher estimated error in $\omega_{\rm cdm}$, $h$ and  $A_s$ as a function of the number of tracers, having $\Delta b_{\rm all} = 0.6$ fixed. For the tracer split with more tracers, we generalize \refeq{deltab2} using evenly spaced MT bias parameters
\be
b_{\O}^{t_i} = b_{\O}^{\rm ST} \pm \Delta b \left(\frac{i-1}{N_{\rm Tracers}-1}-\frac{1}{2}\right)\,,
\ee
with $i$ ranging from $1$ to $N_{\rm Tracers}$. We show results for three different values of $\bar{n}^T = \{10^{-4}, 10^{-3}, 10^{-2}\}  \,h^3{\rm Mpc}^{-3} $.
We find that, for all values of $\bar{n}^T$, $N_{\rm Tracers} = 2$ seems to be the optimal number of tracers, at least when considering only the power spectrum in the analysis. Increasing the number of tracers barely improves the result, making it even slightly worst when  $N_{\rm Tracers} = 5$.  
Finally, we also include the case of more tracers without the cross-stochastic contribution as dotted lines in \reffig{ntracers}. Interestingly, $N_{\rm Tracers} = 2$ also appears to be the optimal case when neglecting cross-stochasticity.

\subsection{The dependence on the total and relative number density} \label{sec:shot}

\begin{figure}[h]
	\centering
	\includegraphics[width = 0.48\textwidth]{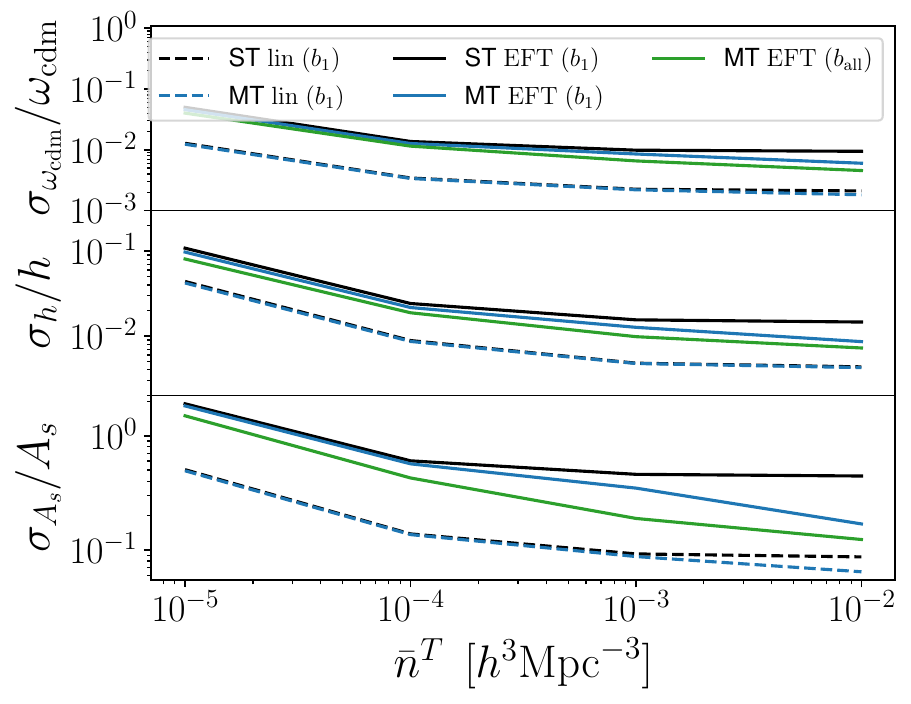}
	\includegraphics[width = 0.48\textwidth]{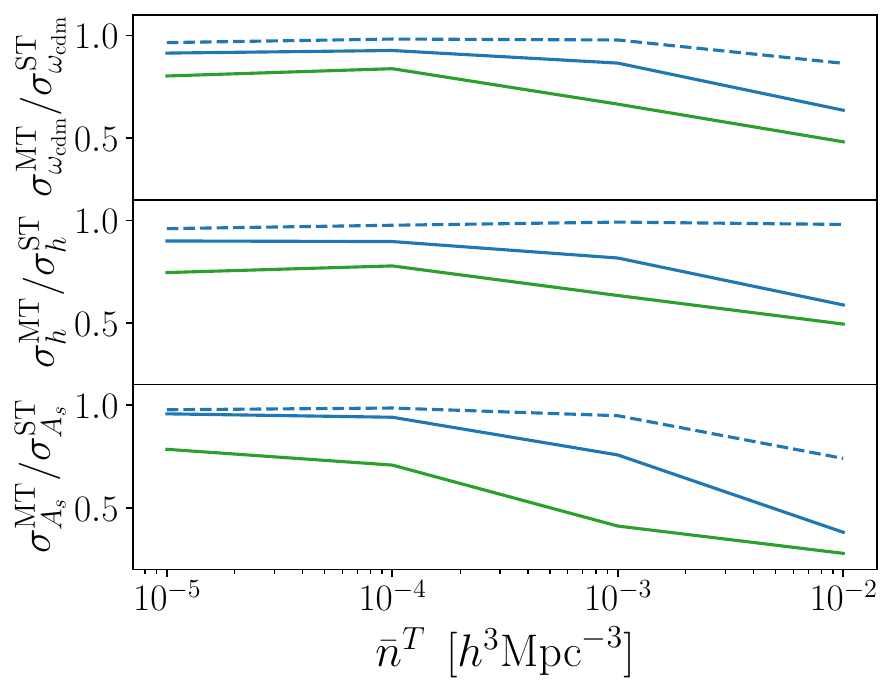}
	\caption{ In the left panel, the relative error in the cosmological parameters as a function of the total number density $\bar{n}^T$. Dashed lines correspond to linear theory while solid lines represent the EFT non-linear theory. For the EFT MT analysis, we consider splits in $b_1$ and $b_{\rm all}$. In the right panel, the same results normalized by the corresponding ST scenarios. We fix $\Delta b_{1} = 0.6$ or $\Delta b_{\rm all} = 0.6$.
	}
	\label{fig:nbar}
\end{figure}

As mentioned, the trace split in MT analysis reduce the relative number density of each subsample increasing their relative shot-noise. Therefore, the MT gains are directly connected to the total tracer density of the initial sample. In this section we investigate the dependence of the MT gain on the total sample density as well as the relative number density of the subsamples. 

We display in the left panel of \reffig{nbar} the relative error as a function of the total number density. We normalize by the ST error in the right panel. The dashed lines represent the linear theory, in which we see that MT barely outperforms ST. As we discussed in \refsec{info}, the success of MT in linear theory strongly depends on neglecting the cross-stochasticity.
The solid lines represent the Fisher errors when including non-linear bias and counter-terms. For the MT non-linear case, we consider both a split in $b_1$ and in $b_{\rm all}$. We can observe that, for the linear case, the main improvement from MT is in $A_s$ and starting at $\bar{n}^{T} > 10^{-3}  \,h^3 {\rm Mpc}^{-3}$. When considering the non-linear case, the MT errors are substantially smaller for a broader range of $\bar{n}^{T}$ and also in $\omega_{\rm cdm}$ and $h$. The split in all the bias parameters together, $b_{\rm all}$, always lead to smaller error bars compared to the usual $b_1$ split. The well known saturation of the Fisher information with $\bar{n}^{T}$ for the ST case happens at both linear and non-linear scenarios. This is not the case for MT both linear and non-linear.

\begin{figure}[h]
	\centering
	\includegraphics[width = 0.7\textwidth]{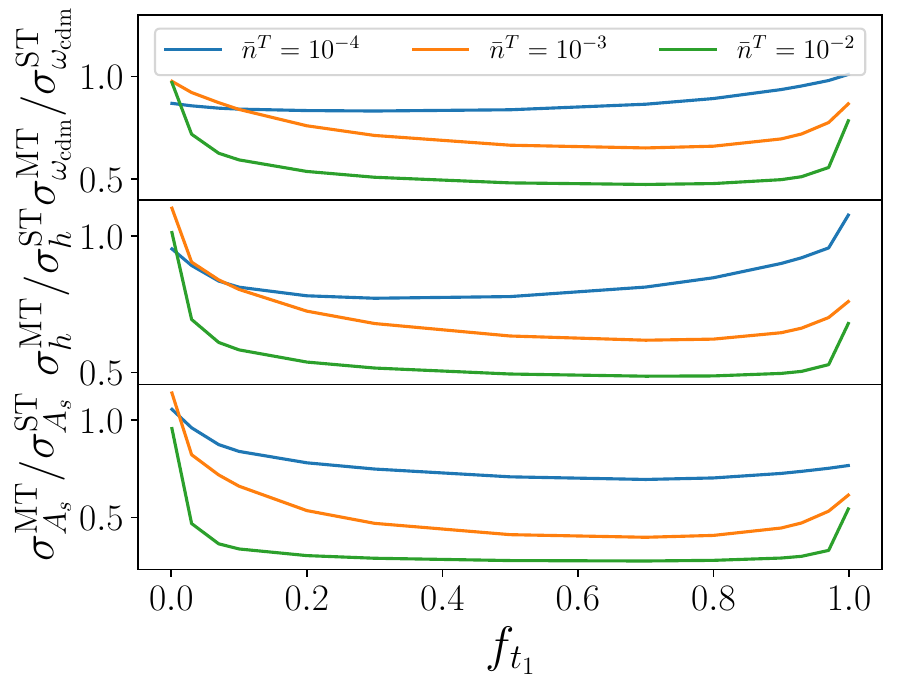}
	\caption{ 
		Error in the cosmological parameters normalized by the ST error as a function of the fraction of the total number density in tracer $t_1$ for a non-balanced split, defined as $f_{t_1} = \bar{n}^{t_1} / \bar{n}^{T}$. Different colors correspond to different values of $\bar{n}^T$, in units of $h^3/$Mpc$^{3}$. As before, we fix $\Delta b_{\rm all} = 0.6$.  
	}
	\label{fig:nsplit}
\end{figure}

When performing a MT split, we also have the freedom to chose the number densities of the final tracers with respect to the initial sample. In \cite{Mergulhao:2021kip,Mergulhao:2023zso} we have only consider a {\it balanced} split, in which each subtracer has the same number density. Here we extend this analysis by studying the MT performance as a function of the tracer $t_1$ fraction, defined in \refeq{frac}. In this work,
we keep the MT relative bias difference $\Delta b$ constant with $f_{t_1}$. However, it is important to note that in realistic scenarios where the two samples are split based on a tracer feature, $\Delta b$ is expected to change with $f_{t_1}$ (see \refsec{toymodel}).
Since this dependence heavily relies on the sample considered, we assume for simplicity a constant bias difference.   
We display in \reffig{nsplit} the error as a function of $f_{t_1}$ for different values of $\bar{n}^{T}$ normalized by the ST error.
Note that the case $f_{t_1} \to 0$ or $f_{t_1} \to 1$, for which the full density is in one of the samples, does not reproduce the ST case for the {\it same} fiducial bias parameters \refeq{fiducialvalues}, but rather corresponds to a different set of bias parameters shifted by $\Delta b$. 
 We see a typical $U$-shape curve with a relatively flat region for $0.1 \lesssim f_{t_1} \lesssim 0.9$, which is especially present as we increase $\bar{n}^T$. It indicates that, despite the optimal value normally being close to the {\it balanced} case $f_{t_1} = 0.5$, the MT improvements are also present with some  unbalancing between both tracer densities. Notice that, in general, it is easier to find subsamples with distinct bias parameters if allowing for their number densities to not be exactly the same. One can, for instance, find the 10 or $20\%$ galaxies with lower or higher values for $b_{\G_2}$ and already benefit from the MT information gains.

\section{FoG and different $k_{\rm max}$} \label{sec:fog}

\begin{figure}[h]
	\centering
	\includegraphics[width = 0.52\textwidth]{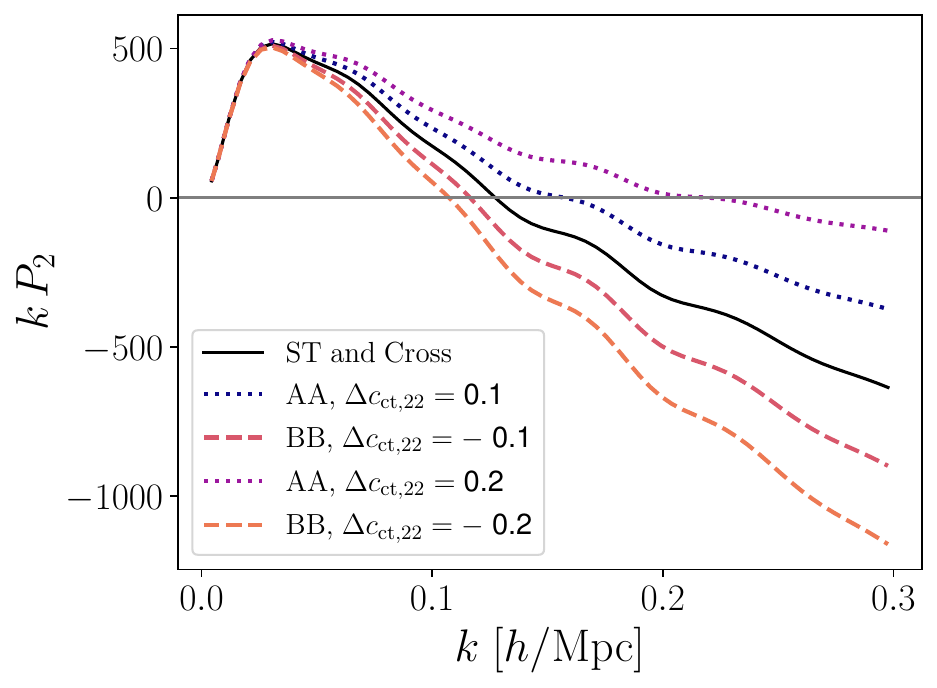}
	\includegraphics[width = 0.46\textwidth]{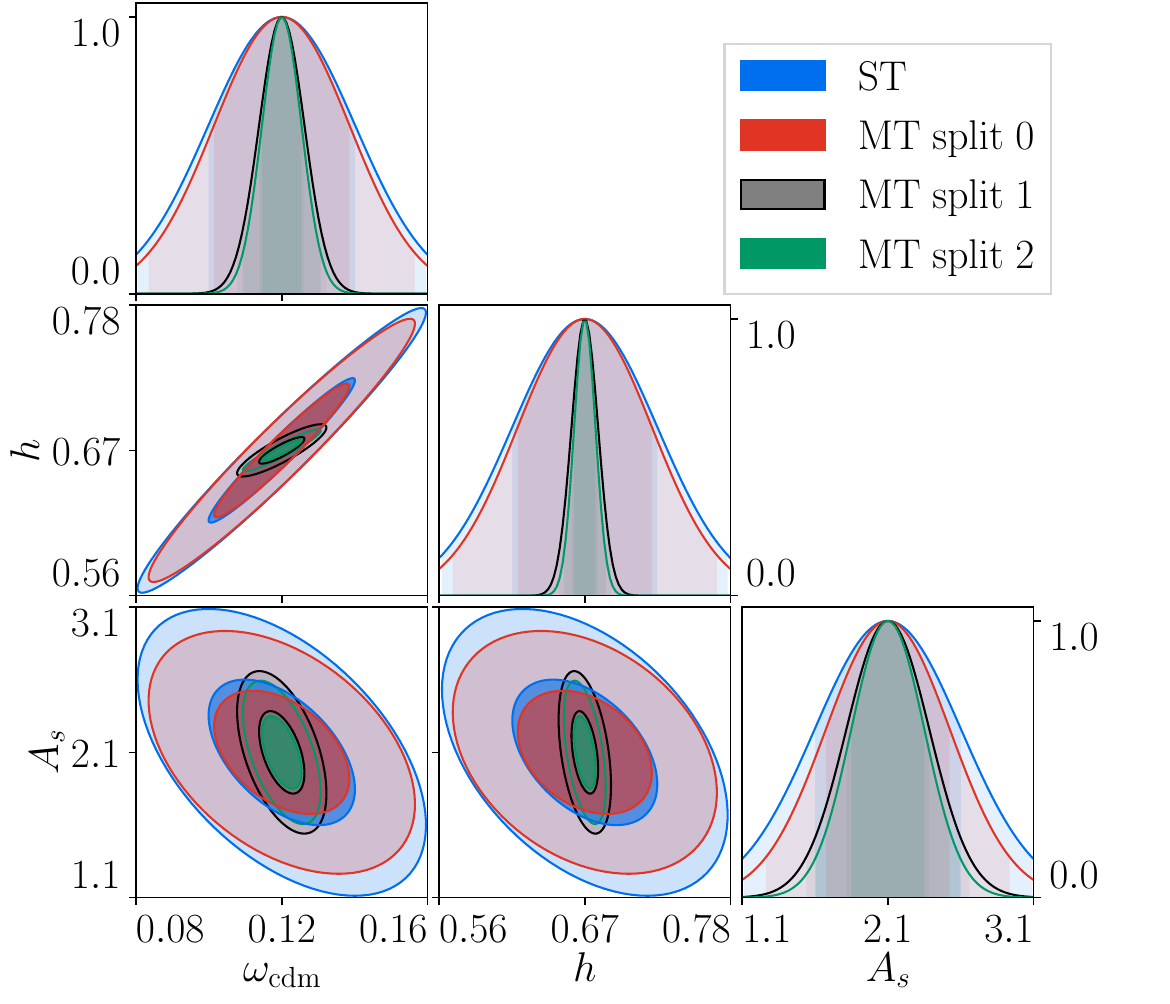}
	\caption{ In the left panel, the quadrupole for the ST case (black), for the blue sample A and for the red sample B. Notice that the cross spectra coincides with the ST line for a tracer split in $c^{\rm ST}_{{\rm ct},22}$. In the right panel, the Fisher contours for different $k_{\rm max}$ cuts in the MT case compared to the ST case (see \reftab{FOGsplit}). }
	\label{fig:P2}
\end{figure}

In this section, we elaborate on how MT can be useful to distinguish between samples with different Fingers-of-God perturbative regimes, such that we could potentially use different $k_{\rm max}$ for different subsamples to maximize the information extracted.
For instance, if one performs a color split, red samples live in more virialized structures and therefore have larger peculiar velocities \cite{Madgwick:2003bd,Coil:2007jp,Hang:2022zyb}. 
The enhancement of the FoG was analysed in the MT context in \cite{Mergulhao:2023zso} (see Fig.~1 therein), where it was shown that the MT split can enhance the FoG effect for one of the samples. 
Improving the FoG treatment can also be done via a rearrangement of the multipole expansion \cite{Ivanov:2021fbu,BOSS:2016off,Kazin:2011xt,DAmico:2021ymi}
or by sub-sample selection \cite{BaleatoLizancos:2025wdg}.
In this work, we consider the possibility of having different scale cuts $k_{\rm max}$ for the two MT subsamples, such that 
we do not throw away any information, but treat the perturbative reach differently for the subsamples. The fact that the bluer subsample suffers less FoG suppression can allow for a sensible increase in its relative $k_{\rm max}$ which leads to information gain. 

The FoG suppression of the quadrupole can be emulated in the bias expansion as a sharp drop in the quadrupole via a large (negative) $c_{{\rm ct}, 22}$ counter-term. To make this effect more pronounced, we adopt in this part the same 
fiducial parameters as \refeq{fiducialvalues} but also taking $c^{\rm ST}_{{\rm ct},22}= -0.3$. 
We perform a MT split in $c_{{\rm ct},22}$ with the idea of mimicking a color split: the red sample is characterized by a more negative $c_{{\rm ct},22}$ and larger FoG suppression, while the blue sample has a more positive $c_{{\rm ct},22}$ value and less FoG suppression on the quadrupole. We illustrate the split in the left panel of  \reffig{P2}. 

Motivated by \cite{BaleatoLizancos:2025wdg}, which treats the zero crossing of the quadrupole a proxy for the perturbative failing of the FoG treatment, we adopt different values of $k_{\rm max}$ for the different samples. We show in \reftab{FOGsplit} the zero-crossing value for the different samples. We notice that a relatively small shift $\Delta c_{{\rm ct},22} = 0.2$ allows for an increase in $k_{\rm max}$ from 0.124$h/$Mpc to 0.208$h/$Mpc. We are assuming that the perturbative scale for dark matter $k_{\rm NL}$ and the higher-derivative expansion scale $1/R_{\rm Halo}$, associated with the Lagrangian halo scale, are smaller than the scales considered here, $k_{\rm NL}>k_{\rm max}$ and $1/R_{\rm Halo}>k_{\rm max}$. We display in the right panel of \reffig{P2} the expected error bars for the different $k_{\rm max}$ considered in \reftab{FOGsplit}. We find a notable improvement compared to the ST case due to the tracer split and the higher $k_{\rm max}$ of the bluer sample.  This result points out to MT as a natural way to improve the perturbative treatment for FoG.    

\begin{table}[!htb]
	\begin{minipage}{1\linewidth}
		\centering
		\begin{tabular}{|c||c|c|c|} \hline 
			& $k_{\rm max}^{AA}$ & $k_{\rm max}^{ST/AB}$  & $k_{\rm max}^{BB}$ \\ \hline \hline
			ST   & -& 0.124& -   \\ \hline
			MT split 0 ($|\Delta c_{{\rm ct}, 22}| = 0.0$)  &  0.124&0.124&0.124 \\ \hline 
			MT split 1 ($|\Delta c_{{\rm ct}, 22}| = 0.1$) &  0.157&0.124& 0.114  \\ \hline 
			MT split 2 ($|\Delta c_{{\rm ct}, 22}| = 0.2$) &  0.208&0.124& 0.105  \\ \hline 
		\end{tabular}
	\end{minipage}%
	\caption{Different $k_{\rm max}$ values (in units of $h/$Mpc) used for the auto, cross and single tracer spectra based on the zero-crossing of the quadrupole.}
	\label{tab:FOGsplit}
\end{table}

\section{Forecasts}
\label{sec:forecasts}

In this section, we forecast the MT scenario for different galaxy survey configurations. For simplicity, we compress the survey data into a single redshift bin. While this is a simplified scenario, it can serve as a good guidance to estimate the MT improvements relative to ST. We use for the effective volume \cite{Chudaykin:2019ock}
\begin{equation}
\label{eq:FKP}
V_{\text{eff}} \approx \sum_i V(z_i) \left[ \frac{ b_1^2(z_i) \Plin(z_i) / \overline{n}(z_i) }{1 + b_1^2(z_i) \Plin(z_i) / \overline{n}(z_i)} \right]^2 _{k = 0.10 \, \frac{h}{\text{Mpc}}}\,,
\end{equation}
summing over multiple redshift bins $z_i$ and anchoring the power spectra at $k = 0.1 \, h/$Mpc. In addition, we compute the effective number density $\overline{n}_{\text{eff}}$ and the effective redshift $z_{\text{eff}}$ by averaging $\overline{n}(z_i)$ and $z_i$ weighted by the number of tracers $N_i = \overline{n}(z_i) V(z_i)$ in the corresponding redshift bin. For the linear bias, we use $b_1^{\rm eff}(z_{\rm eff}) = 0.9+0.4z_{\rm eff}$, which has been used in other works \cite{PFSTeam:2012fqu,Chudaykin:2019ock} and is based on \cite{Orsi:2009mj}. For simplicity, we fix the other non-linear bias parameters to \refeq{fiducialvalues}. We summarize in \reftab{survey} the surveys and their specifications used in this section.

\begin{table}[ht]
	\centering
	\begin{tabular}{|c|ccccc|}
		\hline
		experiment & $z_{\text{eff}}$ & $b_{1, \text{eff}}$ & $\overline{n}_{\text{eff}}$ (in $10^{-3} \frac{h^3}{\text{Mpc}^3}$) & $V_{\rm eff}$ (in $\frac{\text{Gpc}^3}{h^3}$) &  reference\\
		\hline
		BOSS & $0.57$ & $1.13$ & $0.88$ & $2.29$ & \cite{Font-Ribera:2013rwa} \\
		PFS & $1.50$ & $1.50$ & $0.50$ & $2.98$ & \cite{PFSTeam:2012fqu}  \\
		Roman & $1.37$ & $1.45$ & $2.56$ & $8.50$ & \cite{10.1093/mnras/stab1762}\\
		DESI & $1.02$ & $1.31$ & $0.58$ & $12.2$ & \cite{DESI:2016fyo}\\
		Euclid & $0.94$ & $1.27$ & $2.06$ & $24.5$ & \cite{laureijs2011eucliddefinitionstudyreport,Chudaykin:2019ock}\\
		MegaMapper 	& $2.45$ & $1.88$ & $1.69$ & $52.8$ & \cite{Schlegel:2022vrv,Ferraro:2019uce}\\
		\hline
	\end{tabular}
	\caption{Summary of the surveys parameter values considered in this work, along with the references used to obtain their specifications.}
		\label{tab:survey}
\end{table}

\begin{figure}[h]
	\centering
	\includegraphics[width = 0.7\textwidth]{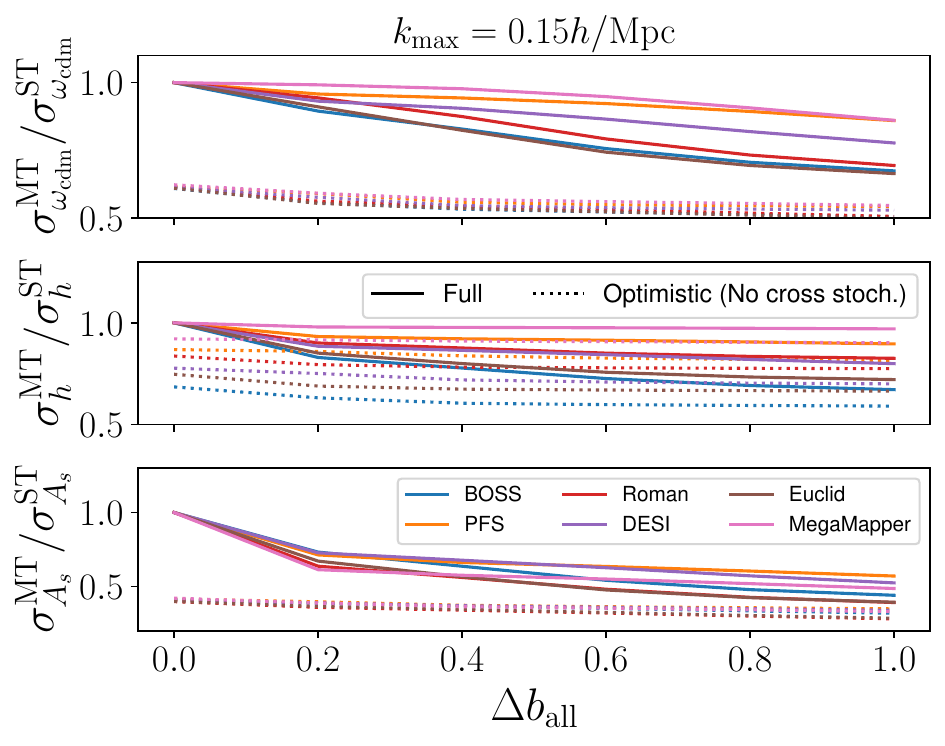}
	\caption{ MT error in the cosmological parameters relative to ST as a function of the difference $\Delta b_{\rm all}$ in all the bias simultaneously. Different colors represent different surveys. The dotted lines represent the analysis without cross-stochasticity.}
	\label{fig:forecast}
\end{figure}

The forecast results are presented in \reffig{forecast} as a function of $\Delta b_{\rm all}$. The solid lines represent the most conservative scenario in which we include all the terms, while the dashed lines represent the most optimistic case in which the cross-stochastic component is neglected. We find a factor 2 improvement in $A_s$ across all surveys. We also find significant improvement in $h$ and $\omega_{\rm cdm}$, but somewhat smaller for PFS and MegaMapper, which we checked to be due the high $z$ considered for those surveys. The non-linear corrections are very small at high redshift at $k = 0.15 \, h/$Mpc, and typically one can reach higher values of $k_{\rm max}$ at high $z$. 
A more comprehensive study which considers the different redshift bins of each survey and models their respective $k_{\rm max}$ values as a function of redshift, is left for future work.
Finally, when the cross-stochastic component is neglected, we find even stronger MT constraints, highlighting the importance of adding simulation-based priors on these parameters.

\section{Conclusions}
\label{sec:conclusion}

In this article, we have extended the results of \cite{Mergulhao:2021kip,Mergulhao:2023zso} carrying a systematic study of MT when considering the non-linear modeling of galaxy clustering. We employ a series of Fisher studies to investigate different scenarios, including various survey configurations.
Within the large-scale bias expansion, any tracer split based on a specific sample feature can be mapped into a direct split in the bias parameters. Accordingly, we display our results as a function of the difference $\Delta b$ in the bias coefficients and show that the MT Fisher information reproduces the ST scenario when taking $\Delta b \to 0$.
Further, we discuss how likely it is to identify two halo samples with differing non-linear bias parameters. While such differences are hard to find in a vanilla scenario, the presence of assembly bias and the use of realistic galaxy samples can lead to $\Delta b_{\G_2} \sim 1$ \cite{Lazeyras:2021dar}{, at least when considering samples with large values of $b_1$.}
This paper paves the way toward identifying a MT bias split that does not necessarily rely on finding two samples with very different {\it linear} bias coefficients. 

Next, we study the importance of cross-stochasticity for MT. {While neglecting this term is vital for the gains of MT {\it within linear theory}, as pointed out by \cite{Gil-Marin:2010ezc,Bernstein:2011ju},} we show that this is not the case when considering the non-linear bias modeling. 
Furthermore, we show for the first time, using non-linear modeling, that a split into two tracers seems to maximize the information gain, at least when restricted to a power spectra analysis.   
We have also found that, when considering the non-linear operators, the MT gains are present at very realistic number density values, in contrast with the findings on linear theory that demands very high tracer densities. 
We addressed the question of what are the optimal number densities between the two subtracers, finding substantial gains even when considering a very unbalanced split, i.e. when one the tracers is more dense than the other. This result makes the task of finding two tracers with two different set of biases simpler, since we can limit the MT searches to find a small in-homogeneous subsample.  
Finally, in \refsec{fog} we consider the possibility of using different values of $k_{\rm max}$ for the subtracers depending on the strength of the FoG effect for each. We show how this approach can substantially enhance the information extracted from the bluer sample.

{We include a summary of the novel results of this work:
\begin{itemize}
    \item We present a systematic study of the multi-tracer approach, incorporating full-shape one-loop corrections. While \cite{Mergulhao:2021kip, Mergulhao:2023zso} focused on specific simulation setups, e.g.~fixed number densities and tracer splits based on halo mass or star formation rate, we explore a wider range of tracer number densities, consider direct splits in the bias parameters and also different noise scenarios.
    \item 
    We find, for the first time, that splitting tracers according to their non-linear bias parameters yields results that are comparable to, and often better than, a simple split in $b_1$. This extends the standard MT framework, which has primarily focused on selecting samples with distinct linear bias, and opens new possibilities for sample selection strategies aimed at maximizing the information extracted from galaxy clustering. We identify assembly bias in the non-linear coefficients as a promising avenue for identifying samples with differing non-linear bias.
    \item 
    We systematically explore the dependence on sample number density for the non-linear case. While, in the linear MT case, improvements only appear at number densities well above $10^{-3} h^3$Mpc$^{-3}$, we find that non-linear MT outperforms ST systematically already at $\bar{n} \sim 10^{-4} h^3$Mpc$^{-3}$.
    \item 
    We investigate, for the first time in the context of non-linear MT, the optimal subsample split as a function of number density, going beyond the $50\%/50\%$ balanced split adopted in \cite{Mergulhao:2021kip, Mergulhao:2023zso}. We find that even a highly unbalanced split, with $90\%/10\%$ of the total number density, can capture a large fraction of the MT information gain, making it considerably easier to identify distinct subsamples.
    \item While the inclusion of cross-stochasticity has traditionally hindered MT performance in the linear full-shape case \cite{Gil-Marin:2010ezc,Bernstein:2011ju}, we find that this is no longer true when non-linear bias terms are included. 
    \item For the first time, we consider more than two tracers in the non-linear modeling of multi-tracer analyses and find that extending beyond two tracers provides only minor gains.
    \item We suggest that a tracer split that leads to different FoG scales, in the lines of \cite{BaleatoLizancos:2025wdg}, can be casted as different $\kmax$ choices for subtracers leading to narrower posteriors in cosmological parameters. 
    \item We forecast the multi-tracer information gain for both current and future survey specifications.
\end{itemize}
}

 We plan to apply in the short future the MT pipeline to (e)BOSS and DESI data. It would also be interesting to extend the PNG analysis of \cite{Barreira:2023rxn} to non-linear scales and investigate the information content of the MT non-linear bispectrum.   Moreover, when splitting into tracers, the running of the bias parameters as a function of the cutoff $\L$ also changes \cite{Rubira:2023vzw}. It would be interesting to investigate the running of the MT bias parameters in the context of the bias renormalization group.

\acknowledgments
We thank Raul Abramo, David Alonso, Yan-Chuan Cai, Anna Cremaschi, Boryana Hadzhiyska, Steffen Hagstotz, Eiichiro Komatsu, Fiona McCarthy, Thiago Mergulh\~ao, Srinivasan Sankarshana, Barbara Sartoris, Fabian Schmidt, Blake Sherwin, Beatriz Tucci and Matteo Zennaro for useful discussions. We thank Alex Barreira, Mathias Garny and Rodrigo Voivodic for valuable comments in the draft.

\appendix
\section{Stability of Fisher derivatives} \label{app:derivatives}

\begin{figure}[h]
	\centering
	\includegraphics[width = 0.49\textwidth]{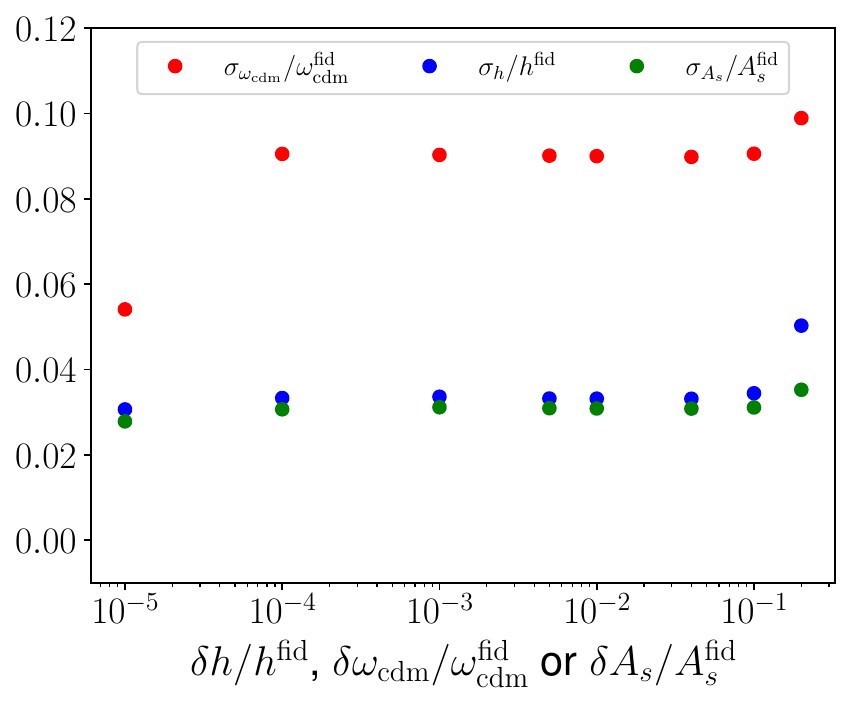}
	\caption{ Relative error $\sigma_h/h^{\rm fid}$, $\sigma_{\omega_{\rm cdm}}/\omega_{\rm cdm}^{\rm fid}$ and $\sigma_{A_s}/A_s^{\rm fid}$ as a function of the relative step size $\delta h$, $\delta \omega_{\rm cdm}$ and $\delta A_s$ used in the (second-order) finite difference method. A small step size is consistent with noise while a large step size leads to the breakdown of the finite difference method. We adopt $\delta h/h^{\rm fid} = \delta \omega_{\rm cdm}/\omega_{\rm cdm}^{\rm fid} = \delta A_s/A_s^{\rm fid} = 0.04$, where the errors remain relatively stable. 
    }
	\label{fig:app_increment}
\end{figure}

The derivatives $\partial_{\theta_a}P^{t_it_j}(k_\alpha)$ in the Fisher analysis of \refeq{FisherMT} can be computed analytically with respect to the bias and stochastic parameters but have to be computed numerically for $h$, $\omega_{\rm cdm}$ and $A_s$. To improve the stability of the numerical derivatives, we use a second-order finite difference method and compute the log-derivative $ P^{t_it_j}(k_\alpha)\partial_{\theta_a}\ln P^{t_it_j}(k_\alpha)$. 
We show in \reffig{app_increment} the stability of the Fisher-extracted relative parameters errors $\sigma_h$, $\sigma_{\omega_{\rm cdm}}$ and $\sigma_{A_s}$ as a function of the step sizes $\delta h$, $\delta \omega_{\rm cdm}$ and $\delta A_s$ used for the finite difference evaluation. 
We observe that choosing a step size that is too small results in numerical noise, while large step sizes lead to the breakdown of the finite difference method. 
Intermediate values lead to relatively stable results. Based on this, we adopt the relative step size $\delta h/h^{\rm fid} = \delta \omega_{\rm cdm}/\omega_{\rm cdm}^{\rm fid} = \delta A_s/A_s^{\rm fid} = 0.04$, normalized by their fiducial Planck 18 values.

\section{Dependence on the fiducial values} \label{app:fiducial}

\begin{figure}[h]
	\centering
	\includegraphics[width = 0.49\textwidth]{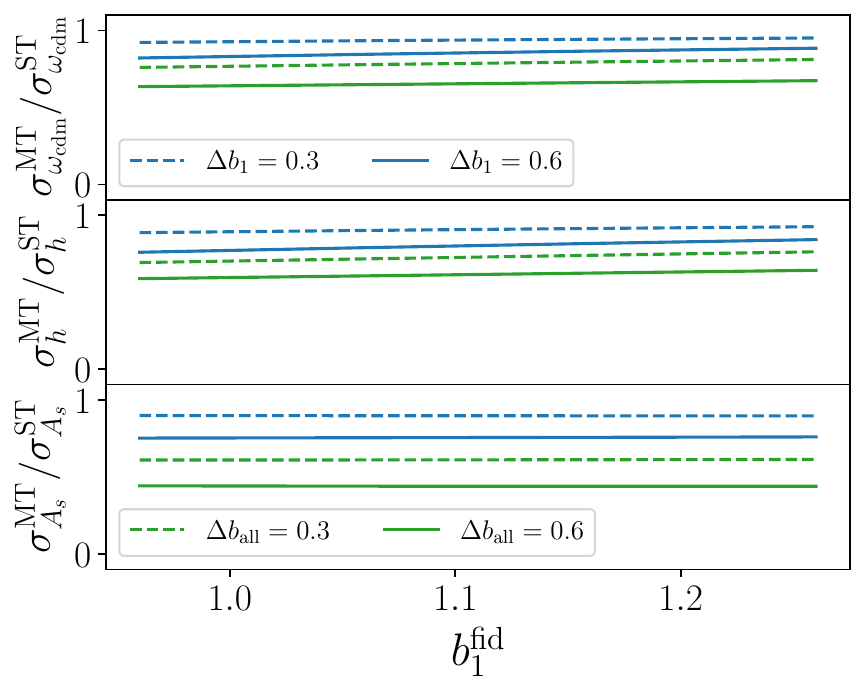}
    \includegraphics[width = 0.49\textwidth]{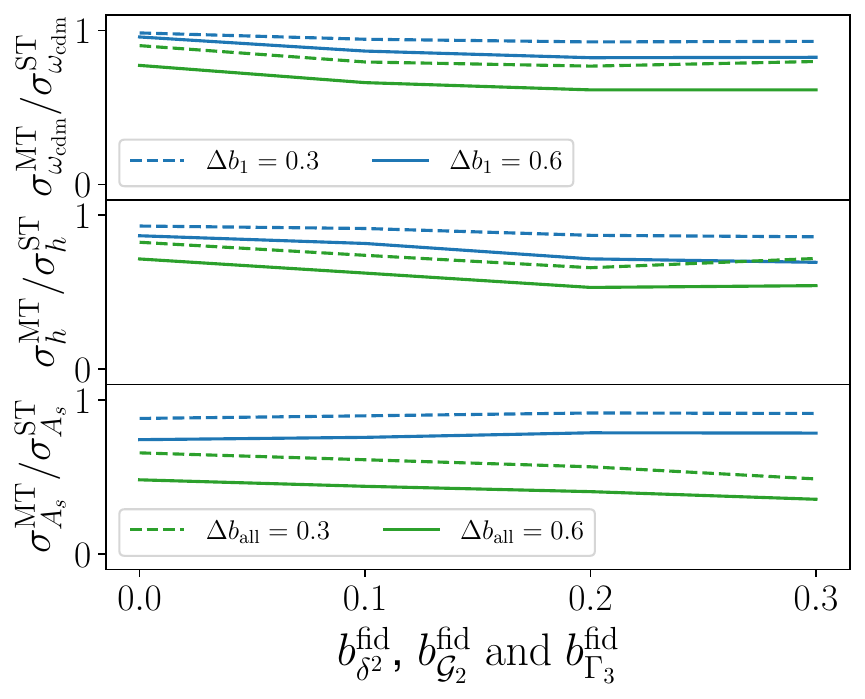}
	\caption{ 
    Ratio between MT and ST error bars as a function of the fiducial bias parameters chosen in \refeq{fiducialvalues}. For the MT results, we consider $\Delta b =0.3$ (dashed) and $\Delta b =0.6$ (solid) splits. {In the left, we vary $b_1^{\rm fid}$ keeping the other parameters fixed to \refeq{fiducialvalues}, in the right we vary the non-linear bias fiducial values, keeping $b_1^{\rm fid} = 1.16$.}}
	\label{fig:fiducial}
\end{figure}

In this section, we discuss how our results depend on the fiducial values chosen for the bias parameters in \refeq{fiducialvalues}. { The left panel of \reffig{fiducial} shows the ratio between the MT and ST error bars as a function of the fiducial values for $b_{1}$}. The right panel of \reffig{fiducial} shows the ratio between the MT and ST error bars as a function of the fiducial values for {$b_{\d^2}$, $b_{\G_2}$ and $b_{\Gamma_3}$, changing all those fiducial bias parameters together}. For the MT results, we considered splits of $\Delta b = 0.3$ (dashed) and $\Delta b = 0.6$ (solid). We observe {very small fluctuations when changing the fiducial value of $b_1$ and fluctuations of at most $20 \%$ when varying the fiducial value of the non-linear bias parameters}, yet the overall qualitative results remain unchanged. 

\section{Extra plots for $b_{\rm all}^{+\pm\pm\pm}$ and different $k_{\rm max}$} \label{app:extraplots}

\begin{figure}[h]
	\centering
	\includegraphics[width = 0.48\textwidth]{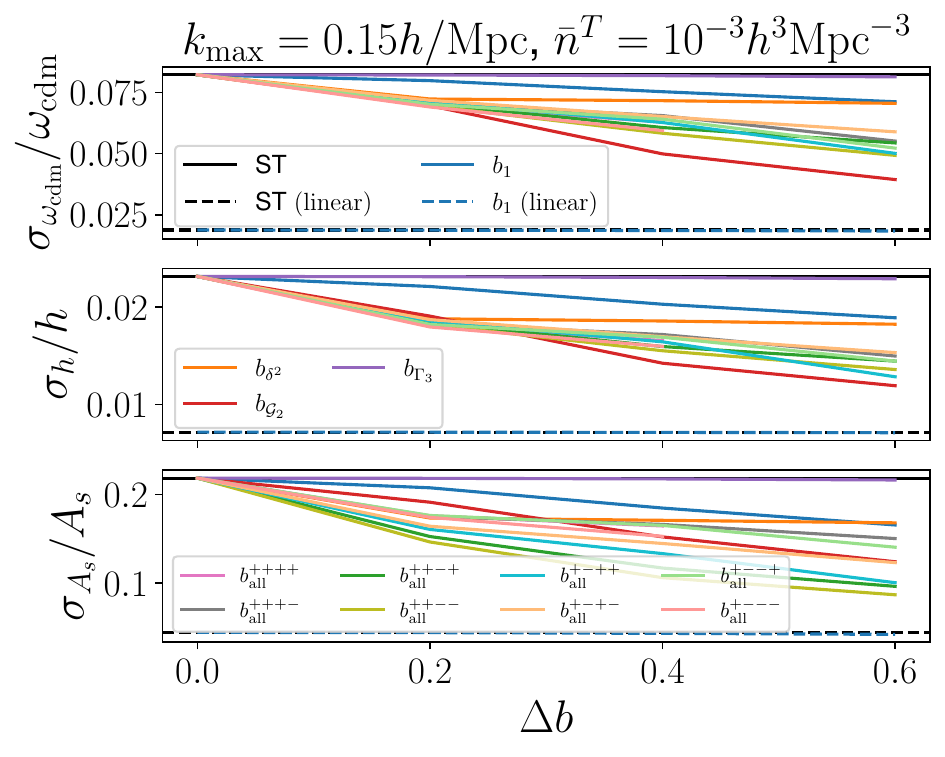}
	\includegraphics[width = 0.48\textwidth]{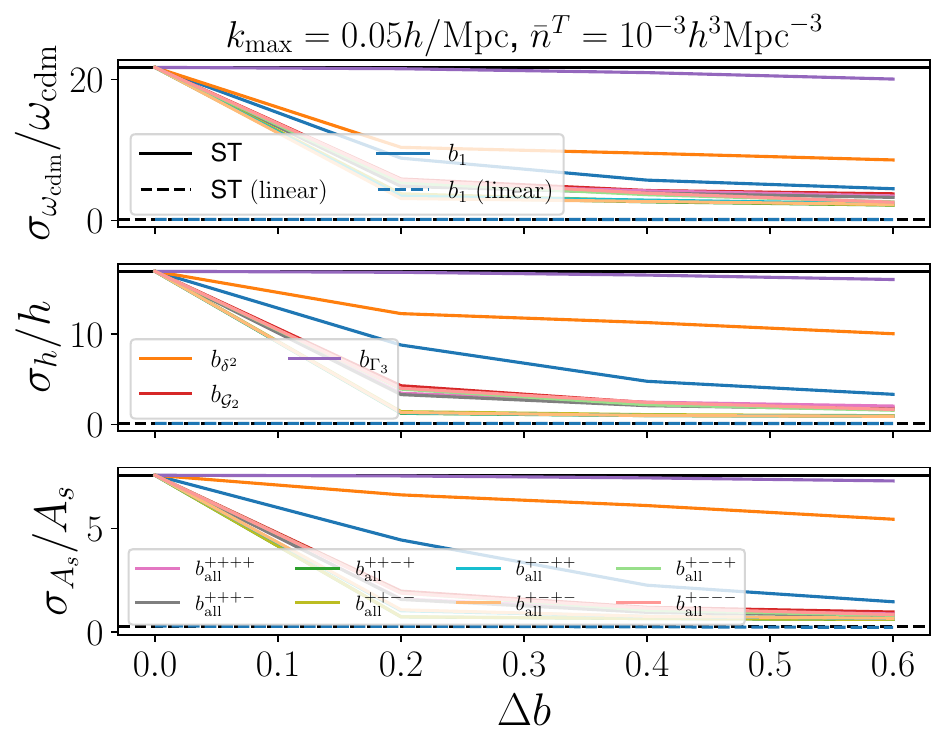}
	\caption{ Same as \reffig{deltab}, but in the left panel we include all combinations of $b_{\rm all}^{\pm\pm\pm\pm}$. The $b_{\rm all}$ lines correspond to a simultaneous split in all four bias parameters, with the $\pm$ sign indicating the sign of $b_1$, $b_{\delta^2}$, $b_{\G_2}$, and $b_{\Gamma_3}$ for the $t_2$ sample, as defined in \refeq{deltab2}. In the right panel, we set $k_{\rm max} = 0.05 h/$Mpc.}
	\label{fig:deltab_all}
\end{figure}

\begin{figure}[h]
	\centering
	\includegraphics[width = 0.6\textwidth]{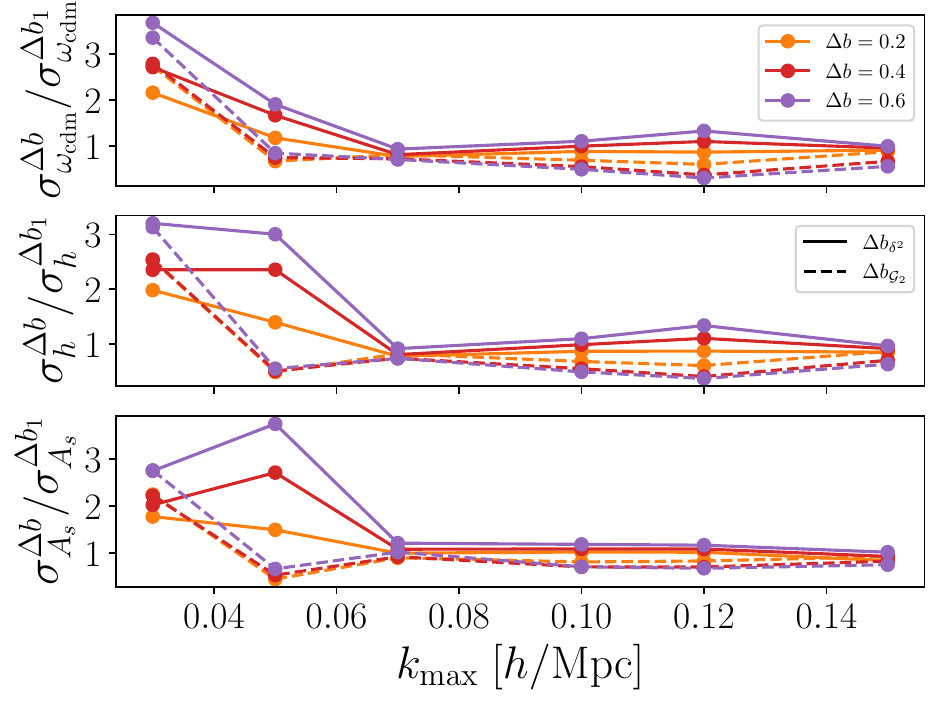}
	\caption{{Error in $\omega_{\rm cdm}$ (top), $h$ (middle) and $A_s$ (bottom) for a split $\Delta b_{\d^2}$ (solid) and $\Delta b_{\G_2}$ (dashed) divided by the error calculated using a split in $\Delta b_{1}$ as a function of $k_{\rm max}$. Different colors represent different $\Delta b$ values. We see that as we consider only linear scales ($k_{\rm max} \to 0$), a split in $b_1$ tends to be more efficient than a split in non-linear bias parameters.} }
	\label{fig:kmax_deltab}
\end{figure}

In this Appendix, we include complementary plots that were not added to the main part of the paper. In \reffig{deltab_all}, the left panel displays all possible combinations of $b_{\rm all}^{+\pm\pm\pm}$; although we observe some relative differences between them, no qualitative change occurs. In the right panel, we present the same plot as in \reffig{deltab}, but with $k_{\rm max} = 0.05\, h/$Mpc. The error bars are, as expected, larger, but we find that a split in $b_1$ becomes relatively more important than a split in $b_{\d^2}$. {We also show in \reffig{kmax_deltab} the error in the cosmological parameters for a split $\Delta b_{\d^2}$ (solid) and $\Delta b_{\G_2}$ (dashed) normalized by the error calculated using $\Delta b_{1}$ as a function of $k_{\rm max}$. As we restric the analysis to large scales, the split in the linear bias parameter becomes more relevant than a split in the non-linear bias. }

\bibliographystyle{JHEP}
\bibliography{main}

\end{document}